\UseRawInputEncoding
\documentclass[%
 reprint,
superscriptaddress,
 amsmath,amssymb,
 aps,
 prl,
]{revtex4-2}

\usepackage{graphicx}
\usepackage{dcolumn}
\usepackage{bm}
\usepackage{xcolor}
\usepackage{ulem}
\usepackage{natbib}
\begin{document}

\preprint{APS/123-QED}

\title{
Small-world complex network generation on a digital quantum processor}

\author{Eric B. Jones}
 \email{Eric.Jones@nrel.gov}
 \affiliation{National Renewable Energy Laboratory, Golden, CO 80401, USA}
\author{Logan E. Hillberry}%
\affiliation{Department of Physics, University of Texas, Austin, TX 78712, USA}


\author{Matthew T. Jones}
\affiliation{Department of Physics, Colorado School of Mines, Golden, CO 80401, USA}%
\affiliation{NVIDIA Corporation, Boulder, CO 80302, USA}

\author{Mina Fasihi}
\affiliation{Department of Physics, Colorado School of Mines, Golden, CO 80401, USA}

\author{Pedram Roushan}
\affiliation{Google Quantum AI, Santa Barbara, CA 93101, USA}

\author{Zhang Jiang}
\affiliation{Google Quantum AI, Santa Barbara, CA 93101, USA}

\author{Alan Ho}
\affiliation{Google Quantum AI, Santa Barbara, CA 93101, USA}

\author{Charles Neill}
\affiliation{Google Quantum AI, Santa Barbara, CA 93101, USA}

\author{Eric Ostby}
\affiliation{Google Quantum AI, Santa Barbara, CA 93101, USA}

\author{Peter Graf}
 \affiliation{National Renewable Energy Laboratory, Golden, CO 80401, USA}

\author{Eliot Kapit}
\email{ekapit@mines.edu}
\affiliation{Quantum Engineering Program, Colorado School of Mines, Golden, CO 80401, USA}
\affiliation{Department of Physics, Colorado School of Mines, Golden, CO 80401, USA}

\author{Lincoln D. Carr}
\email{lcarr@mines.edu}
\affiliation{Quantum Engineering Program, Colorado School of Mines, Golden, CO 80401, USA}
\affiliation{Department of Physics, Colorado School of Mines, Golden, CO 80401, USA}


\date{\today}
\maketitle

\textbf{
Quantum cellular automata (QCA) evolve qubits in a quantum circuit depending only on the states of their neighborhoods \cite{arrighi2019overview, farrelly2020review} and model how rich physical complexity can emerge from a simple set of underlying dynamical rules \cite{bleh2012quantum}. For instance, Goldilocks QCA depending on trade-off principles exhibit non-equilibrating coherent dynamics and generate complex mutual information networks \cite{hillberry2021entangled}, much like the brain \cite{bullmore2009complex}. The inability of classical computers to simulate large quantum systems is a hindrance to understanding the physics of quantum cellular automata, 
but quantum computers offer an ideal simulation platform \cite{nielsen2002quantum, feynman2018simulating}. Here we demonstrate the first experimental realization of QCA on a digital quantum processor, simulating a one-dimensional Goldilocks rule on chains of up to 23 superconducting qubits. Employing low-overhead calibration and error mitigation techniques, we calculate population dynamics and complex network measures indicating the formation of small-world mutual information networks. Unlike random states \cite{arute2019quantum}, these networks decohere at fixed circuit depth independent of system size; the largest of which corresponds to 1,056 two-qubit gates. Such computations may open the door to the employment of QCA in applications like the simulation of strongly-correlated matter \cite{brun2020quantum, duranthon2021coarse, shah2019quantum} or beyond-classical computational demonstrations.}

One of the most profound observations regarding the natural world is that, despite the simple set of physical laws that underpin it, the universe displays a plethora of complex, emergent phenomena, encountered in fields as diverse as biology, sociology, and physics \cite{anderson1972more, anderson2018basic, jensen1998self}. Examples of classical systems where complexity arises as a result of many interacting degrees of freedom are ecosystems, the human brain, and power grids \cite{turcotte2002self}. Certain classical cellular automata (CA) show how complexity can arise from simple rules without the controlling hand of a designer \cite{adamatzky2010game}. CA possess the ability to generate oscillatory, self-replicating structures and in some instances are themselves Turing complete \cite{wolfram1983statistical, wolfram1985cryptography, lindgren1988complexity, chopard1998cellular, cook2004universality}. 

It is known however, that the laws constituting our best model of the universe are quantum mechanical rather than classical \cite{donoghue2014dynamics}. Therefore, in order to simulate the emergence of complexity more fundamentally, one ought to investigate computational models that are predicated upon quantum mechanics.
Goldilocks quantum cellular automata (QCA) \cite{hillberry2021entangled}, are a class of computational models that exhibit emergent complexity despite being constructed from repeated blocks of simple local unitary operators \cite{farrelly2020review}. They involve trade-offs in the local neighborhood such as are known to be sources of complexity in classical systems
and essential to self-organized criticality \cite{carlson2002complexity}. Some Goldilocks QCA have been shown to generate mutual information networks that exhibit signatures of complexity, such as large network clustering, short average path length, and broad node-strength distribution, typically only observed in classical, small-world networks like social or biological networks \cite{hillberry2021entangled}. In addition, QCA have been proposed for other applications such as lattice discretization in the simulation of strongly-correlated matter, quantum field, and gravitational theories \cite{shah2019quantum, brun2020quantum, duranthon2021coarse}. However, the categorical limitation on the ability of classical computers to simulate the time evolution of large quantum systems is a bottleneck for the discovery and exploration of QCA more generally, hampering the theoretical illumination of the class of systems as a whole \cite{nielsen2002quantum}.

\begin{figure*}
\includegraphics[width=.9\linewidth]{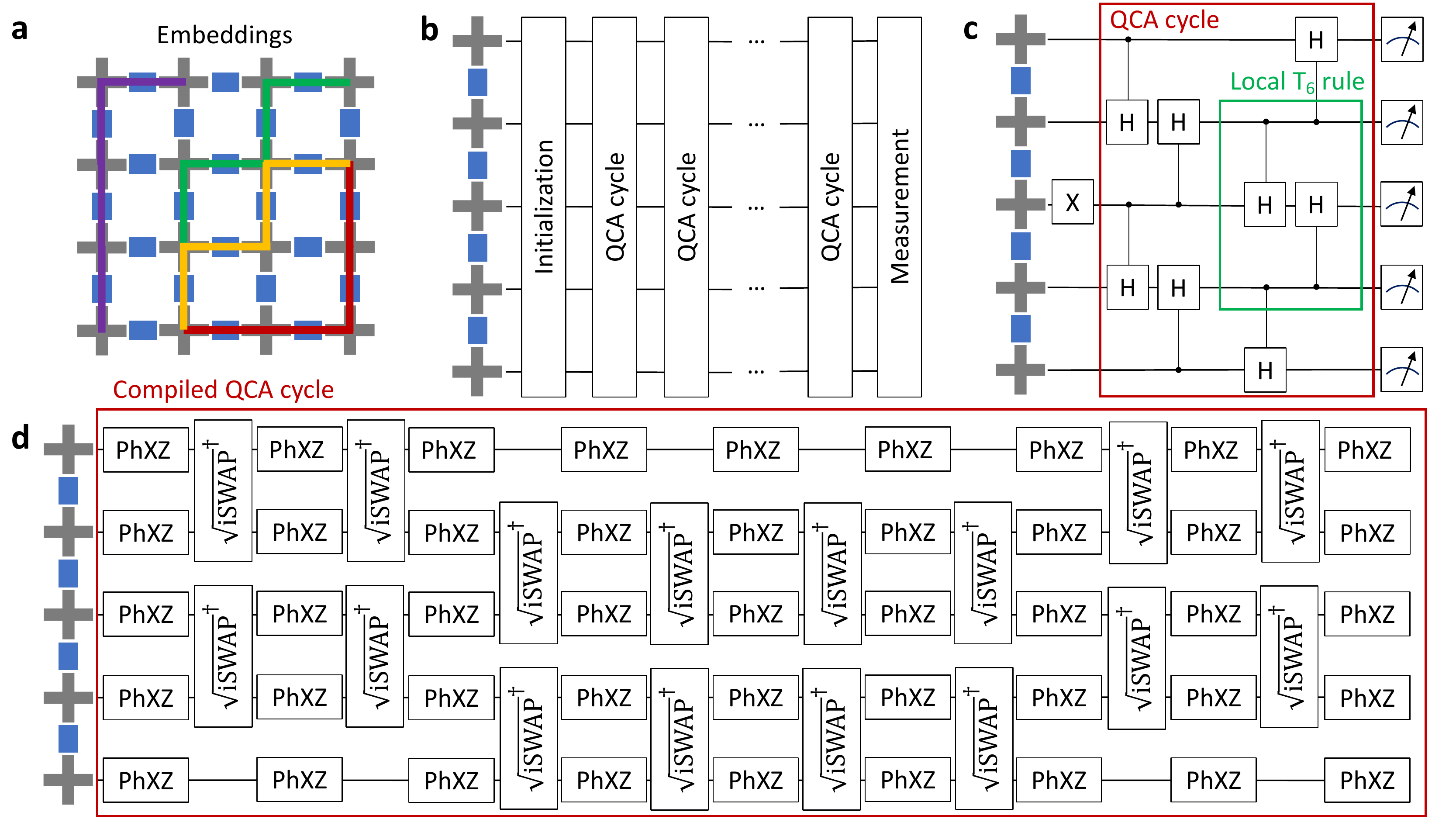}
\caption{\label{fig:circuits} \textbf{One-dimensional quantum cellular automata circuits.} \textbf{a,} Schematic for embedding one-dimensional chains into a subset of a two-dimensional Sycamore-class quantum processor. Grey crosses represent transmon qubits and blue rectangles represent couplers. Purple, green, yellow, and red paths are hypothetical example embeddings. \textbf{b,} Generic structure of a one-dimensional quantum cellular automata (QCA) circuit where time flows to the right. An initialization step is applied to a chain of $L$ qubits, typically to place them into a classical product state with some number of bit flips ($|1\rangle$s). A number of unitary QCA update cycles, $t$, are applied to all $L$ qubits before measurement is performed. \textbf{c,} The specific structure of a Goldilocks QCA for one QCA cycle (red box), wherein the initial state is $|0\ldots 010 \ldots 0\rangle$, the local update unitary is a controlled-controlled-Hadamard gate, and measurement is performed in the computational basis. \textbf{d,} After moment alignment, spin-echo insertion, and compilation down to hardware-native gates a single QCA cycle (red box) results in $4\times(L-1)$ number of $\sqrt{\text{iSWAP}}^{\dagger}$ gates and $8\times L$ number of individually-parameterized $\text{PhXZ}(a, x, z) \equiv \text{Z}^z \text{Z}^a \text{X}^x \text{Z}^{-a}$ gates. 
The number of single and two-qubit layers per QCA cycle does not change as a function of system size, only total gate volume does.
}
\end{figure*}

The last few years have seen the creation of sizeable digital quantum processors that are already demonstrating their value as tools of scientific discovery \cite{wright2019benchmarking, arute2019quantum, arute2020observation, arute2020hartree, pino2021demonstration, neill2021accurately, jurcevic2021demonstration}. Due to their universality, such processors are ideal platforms on which to elucidate the physics and complexity characteristics of QCA. Herein, we simulate a particular one-dimensional QCA on a Sycamore-class superconducting processor depicted schematically in Figs.~\ref{fig:circuits}a-d. Through the calculation of population dynamics and a complex-network characterization of the two-body mutual information matrix we establish that such QCA form small-world mutual information networks and thereby exhibit emergent physical complexity. Moreover, we take the first step towards enabling the widespread use of near-term quantum processors as QCA simulators and offer a template for how to experimentally investigate QCA generally.



\noindent
\textbf{Quantum cellular automata} A one-dimensional (1D) quantum elementary cellular automaton may be defined as a chain of $L$ quantum bits (qubits) whose states are updated according to repeated blocks of neighborhood-local unitary operations along a discrete time axis. When every qubit's state has been updated, a QCA \textit{cycle}, $t$, is completed. After selecting 1D chains of high-quality qubits from the available hardware graph (Fig.~\ref{fig:circuits}a; see also Supplementary Information), the structure (Fig.~\ref{fig:circuits}b) of a 1D QCA experiment is comprised of an initialization step, followed by the application of some number of QCA cycle unitaries out to cycle $t\in \{ 0, 1, \ldots, t_{\text{max}}\}$, and the measurement of appropriate observables. 

\begin{figure*}
\includegraphics[width=.95\linewidth]{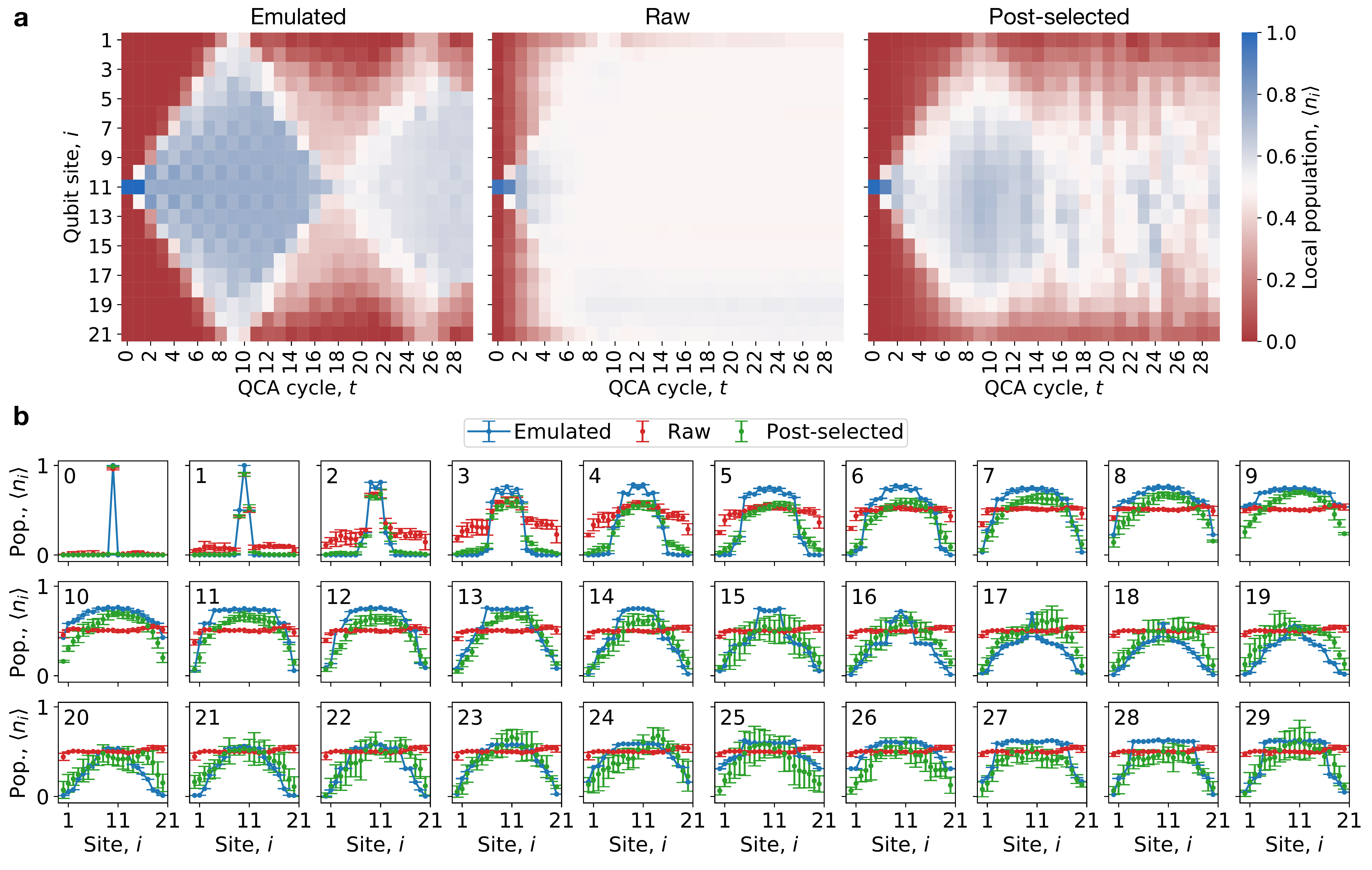}
\caption{\label{fig:populations} \textbf{Population dynamics and post-selection.} \textbf{a,} The left panel shows a noise-free, numerically-emulated Goldilocks QCA out to 30 cycles and initialized with a single $|1\rangle$ in the middle of the 21-qubit chain. Coherent local population, $\langle n_i \rangle$, dynamics that resist equilibration can be observed in the form of the two blue diamond shapes. The middle panel shows raw population dynamics for the same QCA circuit averaged over four 21-qubit chains embedded into the 53-qubit Weber processor. 
The right panel shows the same experimental data but with post-selection applied based on domain wall conservation. 
\textbf{b,} Time-stamped population vignettes show the same dynamics quantitatively for emulated (blue- lines for visual clarity), raw (red), and post-selected (green) data. Error bars represent one standard deviation from the mean over four different chains.
}
\end{figure*}

The particular QCA that we simulate is the totalistic, three-site Goldilocks rule $T_6$ with a uniform Hadamard activation unitary applied to each qubit and boundary conditions equivalent to fixed $|0\rangle$s (see Supplementary Information) \cite{hillberry2021entangled}. We note that the QCA notation $T_6$ should not be confused with decoherence times, which we will denote $\tilde{T}_1$ and $\tilde{T}_2$ where applicable. Fig.~\ref{fig:circuits}c shows how rule $T_6$ is compiled down to quantum gates. A single, central bit flip initialization is followed by one QCA cycle, and finally, measurement in the computational $z$-basis. The local update, represented by two non-Clifford $\text{CH}$ gates (green box), does nothing if there are zero or two adjacent $|1\rangle$s and applies the Hadamard activator to the central qubit if there is exactly one adjacent $|1\rangle$: this is the trade-off rule that gives rise to the \textit{Goldilocks} nomenclature.



\noindent
\textbf{Population dynamics and error mitigation} The quantum processor on which we run our QCA simulations is a 53-qubit superconducting processor, Weber, which follows the design of the Sycamore architecture outlined in Ref. \cite{arute2019quantum} (see also Supplementary Information). Typical performance characteristics for Weber are: single-qubit gate error $e_1\approx 0.1\%$, two-qubit gate error $e_2\approx 1.4\%$, $|0\rangle$-state readout error $e_{r0}\approx 2\%$, $|1\rangle$-state readout error $e_{r1}\approx 7\%$, and population relaxation time $\tilde{T}_1\approx 15 \mu s$ \cite{qcdatasheet}. Fig.~\ref{fig:circuits}d shows the decomposition of a single QCA cycle (red box) to the native $\sqrt{\text{iSWAP}}^{\dagger}$ two-qubit and $\text{PhXZ}(a, x, z) \equiv \text{Z}^z \text{Z}^a \text{X}^x \text{Z}^{-a}$ family of single-qubit gates. Strictly speaking, the native two-qubit gate is better modelled by $\sqrt{\text{iSWAP}}^{\dagger} \times \text{CPHASE}(\varphi)$ where the parasitic cphase is $\varphi \approx \pi/23$ \cite{arute2020observation}. We apply a suite of low-overhead circuit optimization, calibration, and error mitigation techniques to optimize circuit performance including moment alignment, spin-echo insertion, Floquet calibration \cite{arute2020observation, neill2021accurately}, parasitic cphase compensation, and most importantly, post-selection (see Supplementary Information).

\begin{figure*}
\includegraphics[width=1\linewidth]{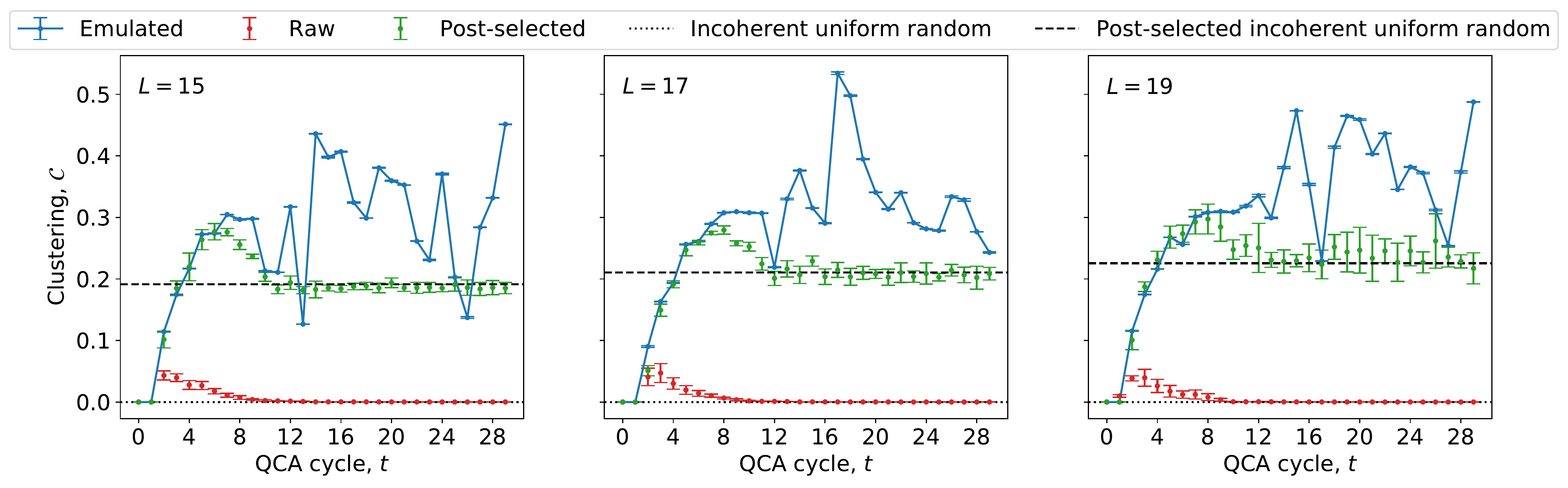}
\caption{\label{fig:clusterings} \textbf{QCA mutual information network clustering.} Clustering coefficient as a function of QCA cycle for three intermediate system sizes simulated on Weber, $L=15, 17, \text{and} \: 19$. Blue curves (lines for visual clarity) are calculated from numerical emulation, while red (green) data points are from Weber data without (with) post-selection. Error bars are one standard deviation in $\mathcal{C}$ over four different chains. Dashed (dotted) black lines are the clustering of the incoherent uniform random state with (without) post-selection.
}
\end{figure*}

At each QCA cycle depth we measure the output of the circuit in the $z$-basis $N_c=100,000$ times, resulting in a set of $L$-bit strings $\{ |z\rangle \}$ and associated probabilities $\{ P_z \approx N_z/N_c \}$, where $N_z$ is the number of times bit string $|z\rangle$ is observed. The local population on each site is calculated via $\langle n_i \rangle = (1-\sum_z P_z (-1)^{z_i})/2$ and averaged over four 1D qubit chains. The left panel of Fig.~\ref{fig:populations}a shows the numerical emulation of such a procedure initialized with a single, central bit flip on 21 qubits, out to 30 QCA cycles. The two large-scale blue diamonds indicate coherent dynamics. When repeated on the Weber processor, a combination of photon loss, gate error, and state preparation and measurement (SPAM) error leads to nearly total population decoherence by $t\approx10$ (Fig.~\ref{fig:populations}a, center panel). We therefore post-select experimental data and discard any measurements whose eigenvalues of the Ising-like operator $\mathcal{O} = \sum_{i=0}^L Z_i Z_{i+1}$ differ from the corresponding eigenvalue of the initial state. That is, $\mathcal{O}$ is a dynamical invariant of the $T_6$ rule that keeps track of the number of domain walls in the system. The right panel of Fig.~\ref{fig:populations}a shows that post-selection results in coherent population dynamics that persist beyond $t\approx 10$, although different observables can degrade with noise on slightly different timescales (see Fig.~\ref{fig:clusterings}). The cycle-stamped population vignettes shown in Fig.~\ref{fig:populations}b support these observations more quantitatively, with error bars representing one standard deviation on the four different chains. After $t \approx 15$, error bars on the post-selected data become more significant and while some qualitative features of the emulated population dynamics appear to persist, such as larger population towards the center of the chain, it is unclear from Fig.~\ref{fig:populations} alone as to what the underlying nature of these qualitative similarities is. Moreover, our complex-network analysis of the behavior of rule $T_6$ relies on the calculation of two-body observables beyond the one-body observables depicted in Fig.~\ref{fig:populations}. As such, we turn to a calculation of Shannon mutual information both to more deeply understand the long-time population dynamics of our QCA and to establish their complex-network behavior.



\begin{figure*}
\includegraphics[width=1\linewidth]{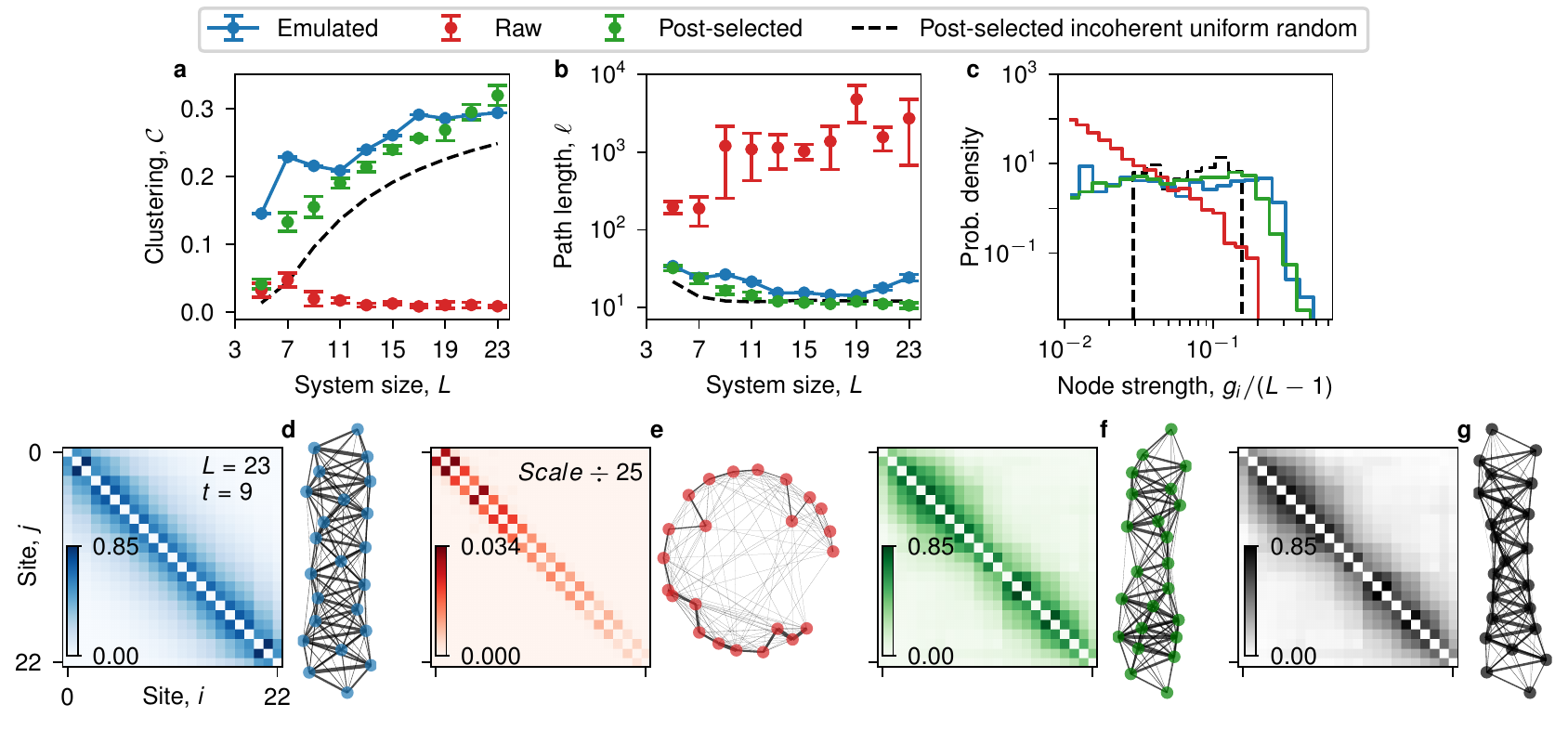}
\caption{\label{fig:complex_network} \textbf{Small-world mutual information network behavior. a,} Coherence window, cycle-and-chain-averaged clustering, $\mathcal{C}$, as a function of system size, $L$. \textbf{b,} Coherence window, cycle-and-chain-averaged path length, $\ell$, as a function of system size, $L$. \textbf{c,} Normalized node strength distribution, $P[g_i/(L-1)]$, amalgamated over all system sizes. \textbf{d-g,} Heatmap and force-directed \cite{fruchterman1991graph} complex network visualization of mutual information network for $L=23$ at $t=9$ (blue: emulated; red: raw data from Weber; green: post-selected; black: post-selected incoherent uniform randomness). Where present, error bars represent one standard deviation from the mean averaged over four different chains and cycles within the coherence window.
}
\end{figure*}

\noindent
\textbf{Mutual information network analysis} Following the complex-network approach in neuroscience wherein functional connectivity of the brain is characterized between spatially non-adjacent regions \cite{bullmore2009complex}, we calculate the classical, Shannon mutual information between all pairs of qubits in each 1D Goldilocks QCA chain

\begin{equation}
    I_{ij} \equiv \sum_{z_i=0}^1 \sum_{z_j=0}^1 p(z_i, z_j) \log_2 \frac{p(z_i, z_j)}{p(z_i)p(z_j)}
\end{equation}
and regard it as an adjacency matrix of correlations that defines the QCA network at each cycle. We choose to use classical, rather than a measure of quantum, mutual information because its calculation only requires measurements in the computational $z$-basis, which we have shown are amenable to post-selection. Moreover, we show in Supplementary Information that the Shannon (classical) mutual information acts as a reliable proxy for von Neumann (quantum) mutual information for the $T_6$ QCA.

Complex networks are ones that are neither purely regular, such as a lattice or complete graph, nor entirely random \cite{albert2002statistical}. The classic demonstration that a network has complex, small-world character involves showing persistently large clustering and simultaneously short path length \cite{watts1998collective}, with a power-law node strength distribution resulting in highly-connected nodes. By analogy with transportation networks, these features describe networks that are easily traversed both locally (clustering) and globally (path length), and exhibit hubs (broad node strength distribution).

Through decoherence, the state of a quantum processor approaches an incoherent uniformly random state with all amplitudes equal to $2^{-L/2}$ and for which $I_{ij} = 0$ for all $L\geq2$. Hence, the incoherent uniformly random state is neither locally nor globally traversable and is thus not a typical random network. Upon post-selection, the decohered state is no longer uniform and the corresponding mutual information network is both non-zero and non-random, although to a much lesser extent than the states generated by Goldilocks QCA. The complexity of networks generated by Goldilocks QCA is established by computing network measures for emulated, raw processor, and post-selected processor states and then comparing each of these to post-selected incoherent uniform random states. 

Clustering measures local network transitivity and is defined as the ratio of the weighted number of closed triangles in the network to the weighted total number of length-2 paths in the network (i.e., the number of \textit{potentially} closed triangles, see Supplementary Information)). The first relevant signature of network complexity is intermediate to large clustering values that do not decay with system size, in contrast to random networks. The emulated clustering (blue curves) of the QCA exhibits this signature and actually increases slightly as a function of system size, $L$ (see Fig.~\ref{fig:clusterings}). While we plot three of the larger system sizes we simulated here, $L=15, 17, \text{and} \: 19$, this proves true for all other system sizes simulated as well. Next, we note that without post-selection, the clustering $\mathcal{C}$ calculated from raw data from Weber (red points) rises briefly but then quickly decays toward zero, the incoherent uniformly random limit (black dotted curve), at $t\approx 12$ for all three system sizes. In contrast, the green curves in Fig.~\ref{fig:clusterings} show that with post-selection the experimental clustering tracks the emulated clustering closely until $t\approx 6$ and remains larger than post-selected uniform randomness (black dashed curve) until $t\approx 12$, independent of system size. There is therefore a window between $t\approx 4$ to $12$ over which we can analyze the formation of a non-random complex network in the QCA for all system sizes simulated. We provide a more detailed description of our cycle windowing process in Supplementary Information.

Fig.~\ref{fig:complex_network}a shows the coherence window, cycle-and-chain-averaged emulated (blue), raw (red), and post-selected (green) clustering coefficient for $L=5$ to $23$ qubits. After the finite-size effects encountered for $L\leq 11$, it is clear that while the raw clustering trends towards zero-- that of a incoherent uniformly random state network-- both the emulated and post-selected clustering stabilize towards $\mathcal{C}\approx 0.3$ and appear to trend towards larger values as a function of system size, indicating substantial network transitivity beyond post-selected randomness (black dashed curve). Fig.~\ref{fig:complex_network}b shows the coherence window, cycle-and-chain-averaged weighted shortest path length, $\ell$, as a function of system size, which gauges global network traversability (see Supplementary Information). The raw data path length (red) in Fig. 4b is large and increases as a function of system size. The post-selected (green) path length tracks the emulated (blue) path length closely, trends downward, and is always one to two orders of magnitude smaller than the raw path length. Interestingly, post-selected path length tracks the path length of post-selected randomness (black dashed line) nearly as well as it does emulated path length. Taken together however, Figs.~\ref{fig:complex_network}a-b signal the existence of small-world mutual information networks generated in the coherence window of the Goldilocks QCA beyond what can be obtained by post-selecting incoherent uniform randomness. The end of the coherence window ($t=12$) for the largest system size simulated ($L=23$) corresponds to $1,056 \: \sqrt{\text{iSWAP}}^{\dagger}$ gates.

Fig.~\ref{fig:complex_network}c further indicates the formation of small-world mutual information networks, showing that the size-normalized emulated (blue) and post-selected (green) node strength distributions (see Supplementary Information), $P[g_i/(L-1)]$, are relatively flat between $1\times 10^{-2}$ to $2 \times 10^{-1}$ compared with those of the post-selected random node strengths (black dashes), which peak between $\sim 2.5 \times 10^{-2}$ to $1.5\times10^{-1}$, and raw (red) node strengths, which are heavily biased towards much smaller values, indicating a deficit in network connectivity. Finally, Figs.~\ref{fig:complex_network}d-g visually depict how the mutual information for $L=23$ at $t=9$ differs for the raw QCA data, which approaches the network structure of incoherent uniform randomness, and the emulated and post-selected QCA networks, which both display lattice girder-like small-world structure that resemble one another more closely than they do post-selected randomness.

\noindent
\textbf{Towards beyond-classical QCA} In addition to their intrinsic scientific value as quantum models for emergent complexity, QCA also present intriguing prospects for establishing new inroads to the beyond-classical era. In the instance of Goldilocks rule $T_6$, identification of a dynamical invariant makes simulation less fragile to noise than a fully chaotic random quantum circuit (RQC). Meanwhile, the system's dynamics still occupy a significant fraction of Hilbert space (see Supplementary Information), which scales as $\sim 1.08^L$.
For context, the long-time, cycle-averaged bond entropy of rule $T_6$ was shown to scale between a 1D area and volume law \cite{hillberry2021entangled}. Although simulation of Goldilocks rules has shown the failure of direct matrix-product-state approaches \cite{vargas2016quantum}, given this intermediate scaling it is an open problem as to whether efficient classical simulatability may be achieved using a modified tensor network approach \cite{hillberry2021entangled, eisert2010colloquium}. Moreover, efficient simulatability of area law-scaling states in two-dimensions (2D) or higher using tensor network approaches, while promising, is even less assured than in 1D.
Hence, 2D QCA that exhibit area-law scaling (or worse) may be good candidates for beyond-classical demonstrations.

Here we have demonstrated that existing quantum processors can efficiently simulate 1D QCA with high fidelity at large gate volume. While reliant on the availability of high-fidelity hardware, the main circuit design principles that enable this goal consist of identification of particular QCA rules that: i) generate significant complexity signatures, ii) efficiently compile to low-depth sequences of hardware-native gates, and iii) are amenable to post-selection through identification of one or more dynamical invariants. We therefore expect these design principles to aid in discovering QCA that support beyond-classical demonstrations or are otherwise useful in quantum computational domain applications. In particular, employing such design principles for QCAs that model correlated quantum matter could be a promising route toward beyond-classical simulation of novel physical systems in the near term.

\bigskip

\noindent
\textbf{Methods} Please see Supplementary Information.

\noindent
\textbf{Data availability} The data supporting this work will be made available upon reasonable request to the corresponding authors.

\noindent
\textbf{Code availability} The code supporting this work will be made available upon reasonable request to the corresponding authors.

\noindent
\textbf{Acknowledgements} 
We thank the Google Quantum AI team. This work was supported in part by the NSF under grant PHY-1653820 (EK); and DGE-2125899, CCF-1839232, PHY-1806372, and OAC-1740130 (MF, MTJ, LDC). This work was authored in part by the National Renewable Energy Laboratory (NREL), operated by Alliance for Sustainable Energy, LLC, for the U.S. Department of Energy (DOE) under Contract No. DE-AC36-08GO28308. This work was supported in part by the Laboratory Directed Research and Development (LDRD) Program at NREL. The views expressed in the article do not necessarily represent the views of the DOE or the U.S. Government. The U.S. Government retains, and the publisher, by accepting the article for publication, acknowledges that the U.S. Government retains, a nonexclusive, paid-up, irrevocable, worldwide license to publish or reproduce the published form of this work, or allow others to do so, for U.S. Government purposes.

\noindent
\textbf{Author contributions} 
Conceptualization (EK, LDC).  
Data curation (EBJ, LEH, MTJ, MF, ZJ, AH).
Formal Analysis (EBJ, LEH, MTJ, MF, PR, ZJ, AH, CN, EO, EK, LDC).
Funding acquisition (PG, EK, LDC).
Investigation (EBJ, LEH, MTJ, MF, PR, ZJ, AH, CN, EO, EK, LDC).
Methodology (EBJ, LEH, MTJ, MF, PR, ZJ, AH, CN, EO, EK, LDC).
Project administration (PR, AH, EO, EK, LDC).
Resources (PR, ZJ, AH, CN, EO).
Software (EBJ, LEH, MTJ, ZJ, AH).
Supervision (PG, EK, LDC).
Validation (all authors).
Visualization (EBJ, LEH, PR, ZJ, AH, CN, EK, LDC).
Writing – original draft (EBJ).
Writing – review \& editing (all authors).

\noindent
\textbf{Competing interests} The authors declare no competing interests.

\noindent
\textbf{Materials \& Correspondence} Materials requests to E.B. Jones. Correspondence to E.B. Jones, E. Kapit, or L.D. Carr.


\newpage

\part{Supplementary Information}

\tableofcontents

\section{Quantum Cellular Automata in 1D}

A one-dimensional quantum cellular automaton is defined as a chain of $L$ identical qubits whose states are updated according to homogeneous blocks of neighborhood-local unitary operators \cite{arrighi2019overview, farrelly2020review}. Each local unitary takes as an input the state of the target qubit's neighbors and outputs a particular operator, either the identity or the activation operator, to be applied to the target qubit, conditioned on the neighborhood's state. When such an update unitary has been applied to all $L$ qubits, a QCA \textit{cycle} is complete. In the specific construction with nearest-neighbor connectivity, corresponding to three-site (closed) neighborhoods (denoted $T_R$), a specific qubit's open neighborhood (its two neighbors) can be in a superposition of any $2^2$ states. The target qubit's state can either be activated or not depending upon these $4$ configurations. As such there are $2^{2^2}=16$ possible transition functions on a three-site closed neighborhood. We label each of these transition functions, or \textit{rules}, as $R\in \{0, 1, \ldots, 15\}$. To specify a three-site rule, we write $T_R$.  The rule which activates for only a balanced neighborhood of 0-1 or 1-0, $T_6$, is called the \textit{Goldilocks rule} and is the main focus of our study on the quantum processor.

For efficient circuit parallelization purposes, we update all even qubits in parallel followed by all odd qubits. In this ordering, the general form for the $L$-qubit unitary operator that evolves the system from cycle $t$ to cycle $t+1$ is \cite{hillberry2021entangled}
\begin{equation}\label{eq:qca_unitary}
    \begin{split}
    U(T_R;t, t+&1) = \prod_{o=1, 3, \ldots}^{L} U_o(T_R)
    \prod_{e=2, 4,\ldots}^{L-1} U_e(T_R), \\
    U_{e(o)}(T_R) = &\sum_{m, n=0}^1 P_{e(o)-1}^{(m)} \otimes V^{c_{mn}}_{e(o)} \otimes P_{e(o)+1}^{(n)}
    \end{split}
\end{equation}
where $i=e(o)$ indexes each even (odd) qubit in the 1D chain, $P_j^{(m)} = |m\rangle \langle m|$ is a single-qubit projection operator, and $V_i$ is the chosen activation unitary applied to qubit $i$ (which is typically taken to be the same operator for all $i$). In order to convert from a rule number to a unitary operator, one first performs a binary expansion of the rule number where the expansion coefficients are themselves labelled by 2-bit strings, $R=\sum_{m,n=0}^1 c_{mn}2^{m+n}$. The coefficients $c_{mn}$ describe which neighborhood configurations activation occurs on. For Goldilocks rule $T_6$ considered in the main text, we have that $6=1\times 2^1 + 1\times 2^2$ so that $c_{00}=c_{11}=0$ while $c_{01}=c_{10}=1$. Therefore, the local $T_6$ cycle unitary reads
\begin{equation}
    U_i(T_6)=|00 \rangle 1\langle 00| + |01 \rangle V\langle 01| +|10 \rangle V\langle 10| + |11 \rangle 1\langle 11|,
\end{equation}
which applies the activation unitary $V$ to qubit $i$ if its surrounding qubits are in the configurations $|01\rangle$ or $|10\rangle$ and does nothing otherwise. We refer to $T_6$ as a \textit{totalistic} rule because activation obeys a left-right symmetry and only depends on the total number of adjacent $|1\rangle$s. Another important totalistic rule is $T_1$, also called the PXP model in many-body quantum literature when run in continuous time~\cite{hillberry2021entangled}; here we consider discrete time, i.e., a quantum circuit, in line with the main concept of cellular automata.  Note that that for 1D, three-site neighborhoods there are only $2^3=8$ totalistic rules. Restricting a QCA search space to require totalistic updates is a useful tool for discovering emergent complexity. Finally, we note that $V=X$, the bit-flip operator, corresponds to a reversible classical cellular automaton. In the main text we choose $V=H$, the Hadamard operator. It has been shown that the particular form of the activation unitary affects the complexity outcomes of $T_6$ relatively little so long as $V$ is able to move classical states off the poles of the Bloch sphere, which $H$ does \cite{hillberry2021entangled}.

\section{Quantum Hardware Specifications and Qubit Picking}

\subsection{Weber Processor}

\begin{figure}
\includegraphics[width=1.\linewidth]{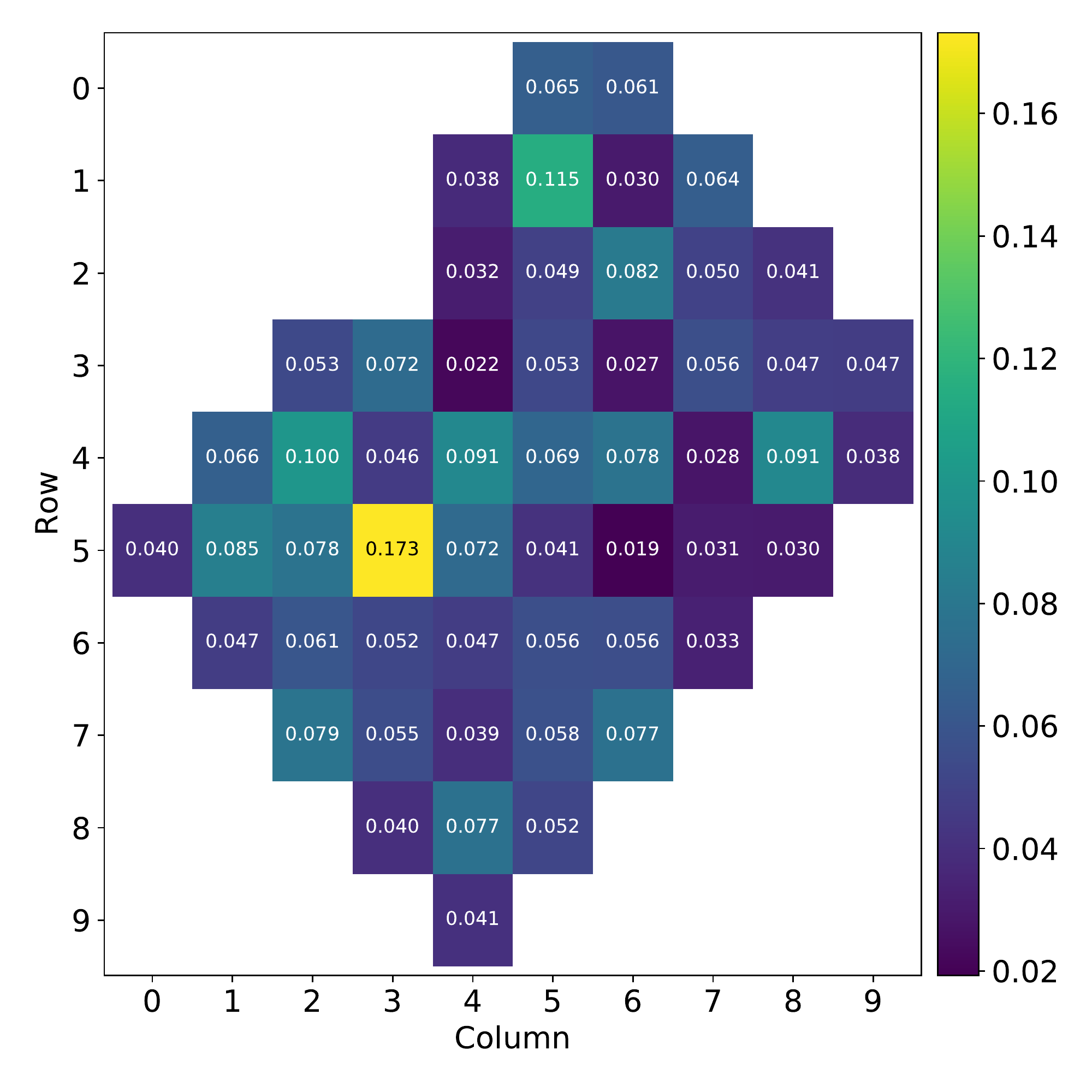}
\caption{\label{fig:readout_error} \textbf{Weber parallel $|1\rangle$-state readout error rates.} Each qubit is represented by a square and is indexed by a row and column number. Heatmap values represent the probability of finding a qubit in the $|0\rangle$ state after it was prepared in the $|1\rangle$ state when all qubits are read out in parallel.
}
\end{figure}

\begin{figure}
\includegraphics[width=1.\linewidth]{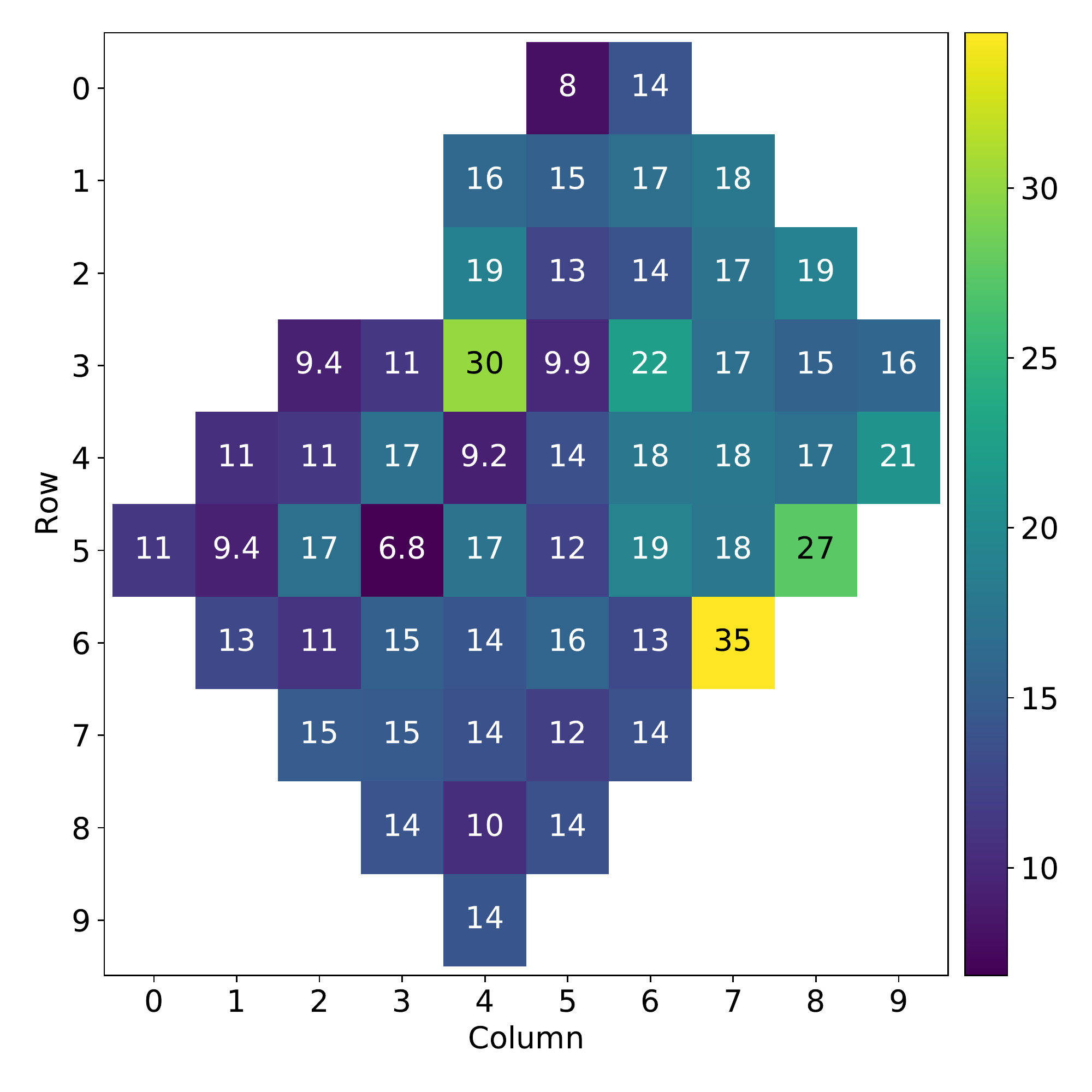}
\caption{\label{fig:t1_time} \textbf{Weber idle $\tilde{T}_1$ decoherence times ($\mu$s).} Each qubit is represented by a square and is indexed by a row and column number. Heatmap values represent the idle $\tilde{T}_1$ decoherence time in $\mu$s.
}
\end{figure}

\begin{figure}
\includegraphics[width=1.\linewidth]{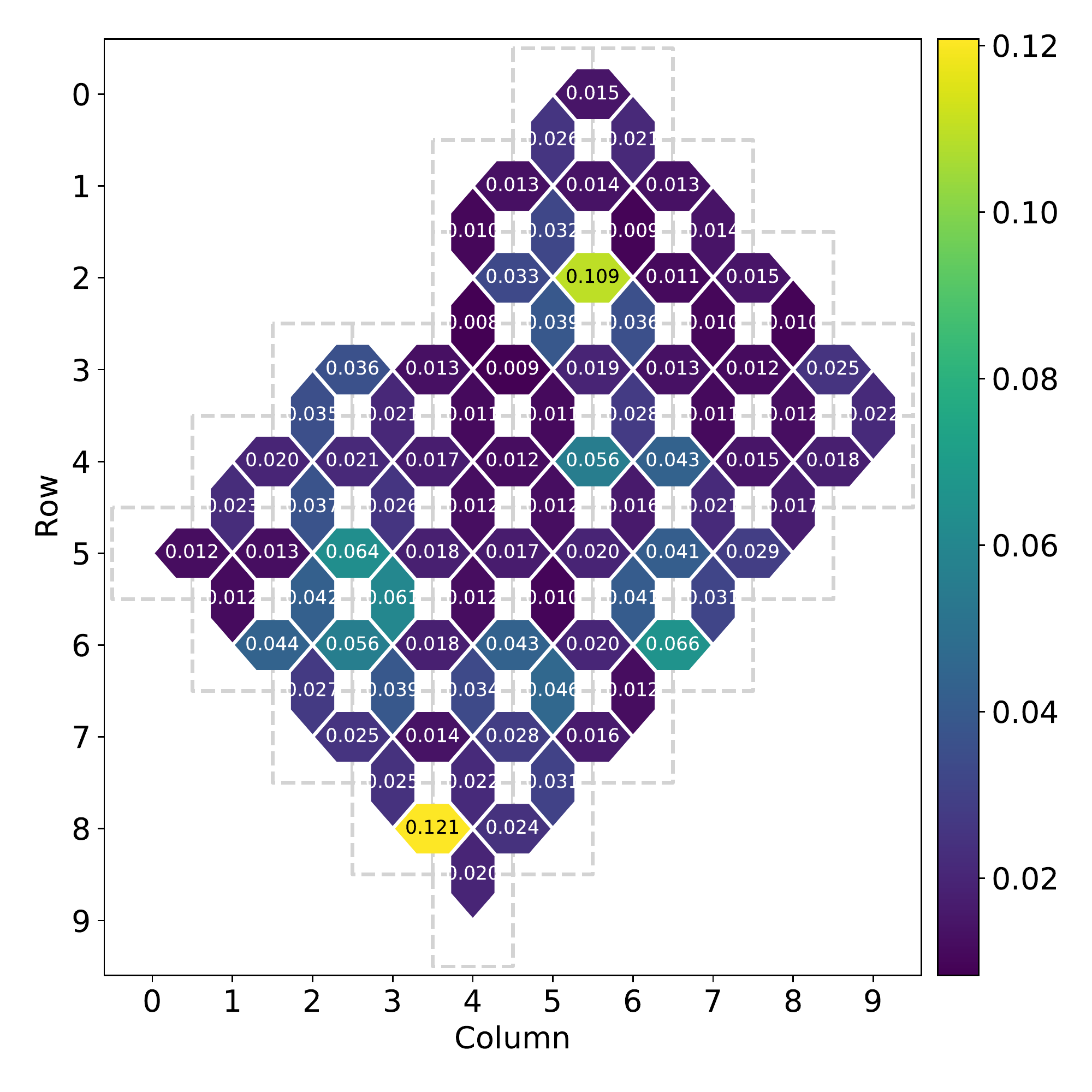}
\caption{\label{fig:two_qubit_error} \textbf{Weber parallel $\sqrt{\text{iSWAP}}^{\dagger}$ Pauli error rates.} Each qubit is indexed by a row and column number, and hexagons represent couplers. Heatmap values represent the cross-entropy benchmarking Pauli error per cycle when executed in parallel (see \cite{arute2019quantum}).
}
\end{figure}

The Weber quantum processing unit is a 53-qubit superconducting processor that follows the design of the Sycamore-class architecture in Ref.~\cite{arute2019quantum}. As discussed in the main text, typical performance characteristics for Weber are: single-qubit gate error $e_1\approx 0.1\%$, two-qubit gate error $e_2\approx 1.4\%$, $|0\rangle$-state readout error $e_{r0}\approx 2\%$, $|1\rangle$-state readout error $e_{r1}\approx 7\%$, and population relaxation time $\tilde{T}_1\approx 15 \mu s$ \cite{qcdatasheet}. Of these, the three performance characteristics that are potentially the most deleterious to calculating accurate observables are $e_{r1}$, $e_2$, and $\tilde{T}_1$ (relative to total circuit execution time). Selecting high-quality chains of qubits therefore involves simultaneously optimizing for these three characteristics, which change slightly between processor calibrations.

For the particular calibration displayed, Fig.~\ref{fig:readout_error} shows that while the parallel readout $e_{r1}$ on Weber is relatively homogeneous across the chip, it is slightly lower on the right-hand side. Meanwhile, the idle $\tilde{T}_1$ decoherence times shown in Fig.~\ref{fig:t1_time} are somewhat longer on the right-hand side of the processor as well and relatively uniform within its top right quadrant. Finally, $e_2$ rates for the $\sqrt{\text{iSWAP}}^{\dagger}$ gate when executed in parallel on Weber and characterized by cross-entropy benchmarking (XEB) are lower both in the processor's top-right quadrant and in a small region left of center. Taken together, these observations indicate that the top-right quadrant of Weber is an ideal region within which to choose 1D qubit embeddings. For instance, a particular five qubit embedding that exploits this quadrant has qubit indices (1, 6) $\rightarrow$ (2, 6) $\rightarrow$ (2, 7) $\rightarrow$ (3, 7) $\rightarrow$ (4, 7). Of course, as chains become longer, that is, as $L$ becomes larger, the available real-estate for picking the highest-quality qubits becomes constrained.

\subsection{Rainbow Processor}

\begin{figure}
\includegraphics[width=1.\linewidth]{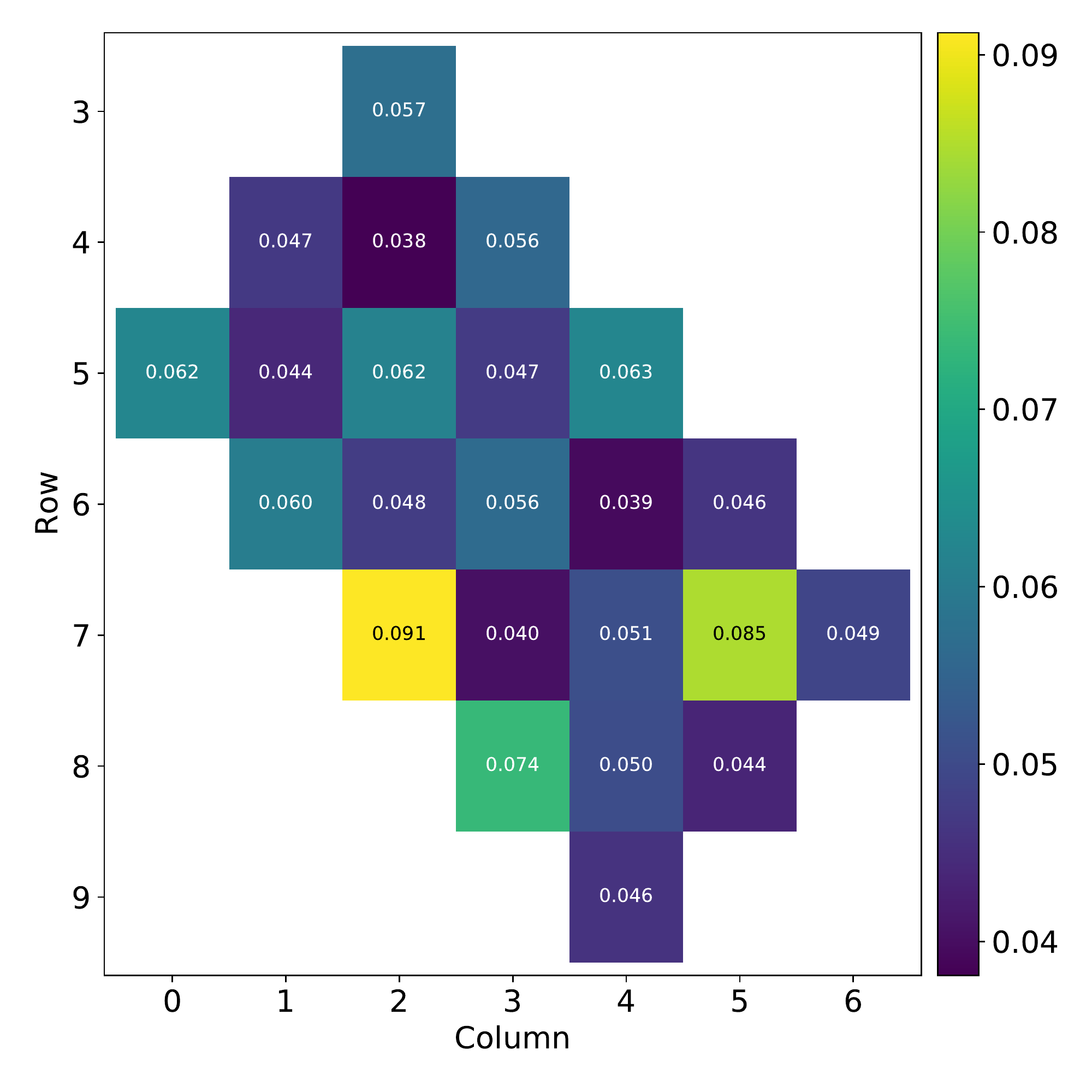}
\caption{\label{fig:readout_error_rain} \textbf{Rainbow parallel $|1\rangle$-state readout error rates.} Same data as Fig.~\ref{fig:readout_error}, but for the 23-qubit Rainbow quantum processor.
}
\end{figure}

\begin{figure}
\includegraphics[width=1.\linewidth]{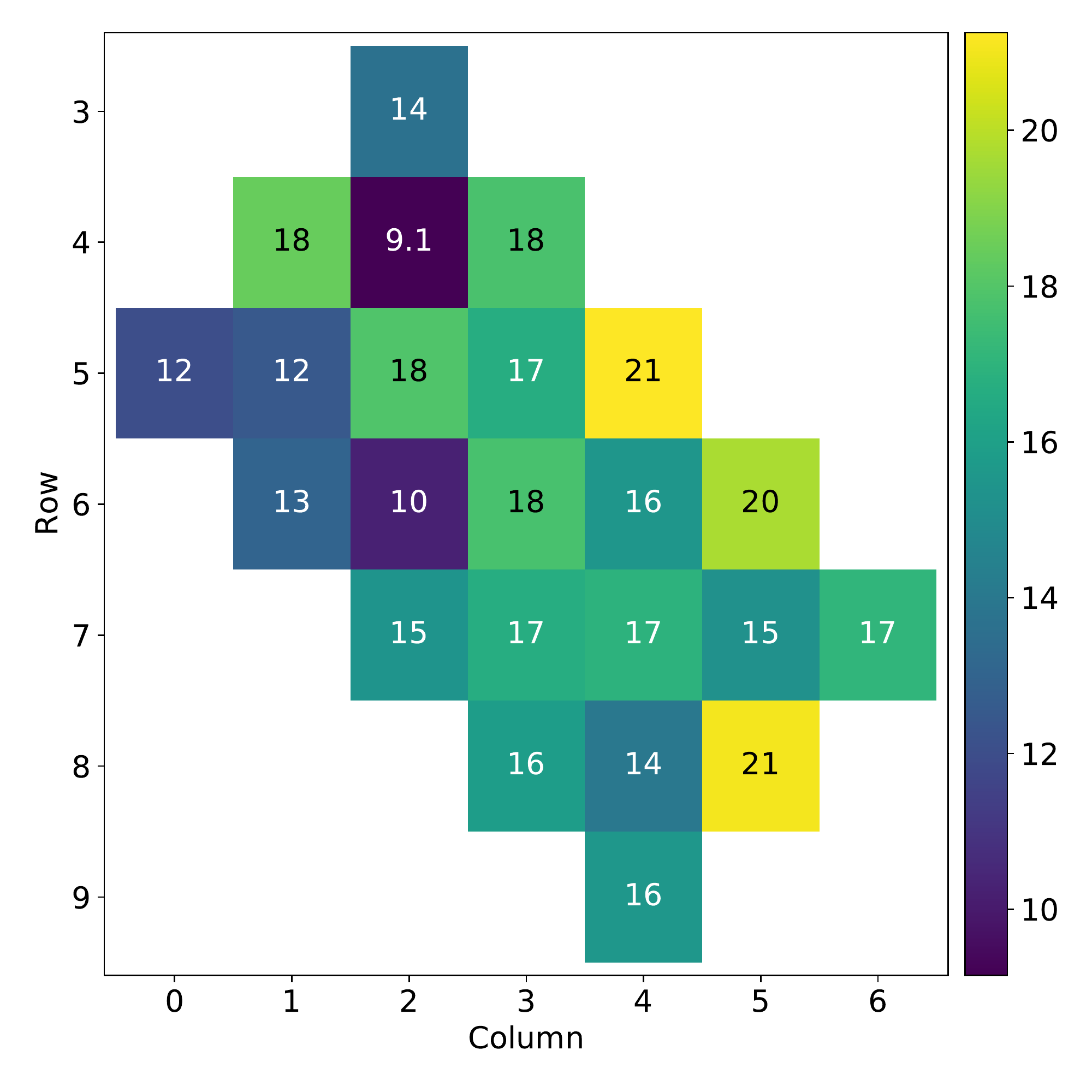}
\caption{\label{fig:t1_time_rain} \textbf{Rainbow idle $\tilde{T}_1$ decoherence times.} Same data as Fig.~\ref{fig:t1_time}, but for the 23-qubit Rainbow quantum processor.
}
\end{figure}

\begin{figure}
\includegraphics[width=1.\linewidth]{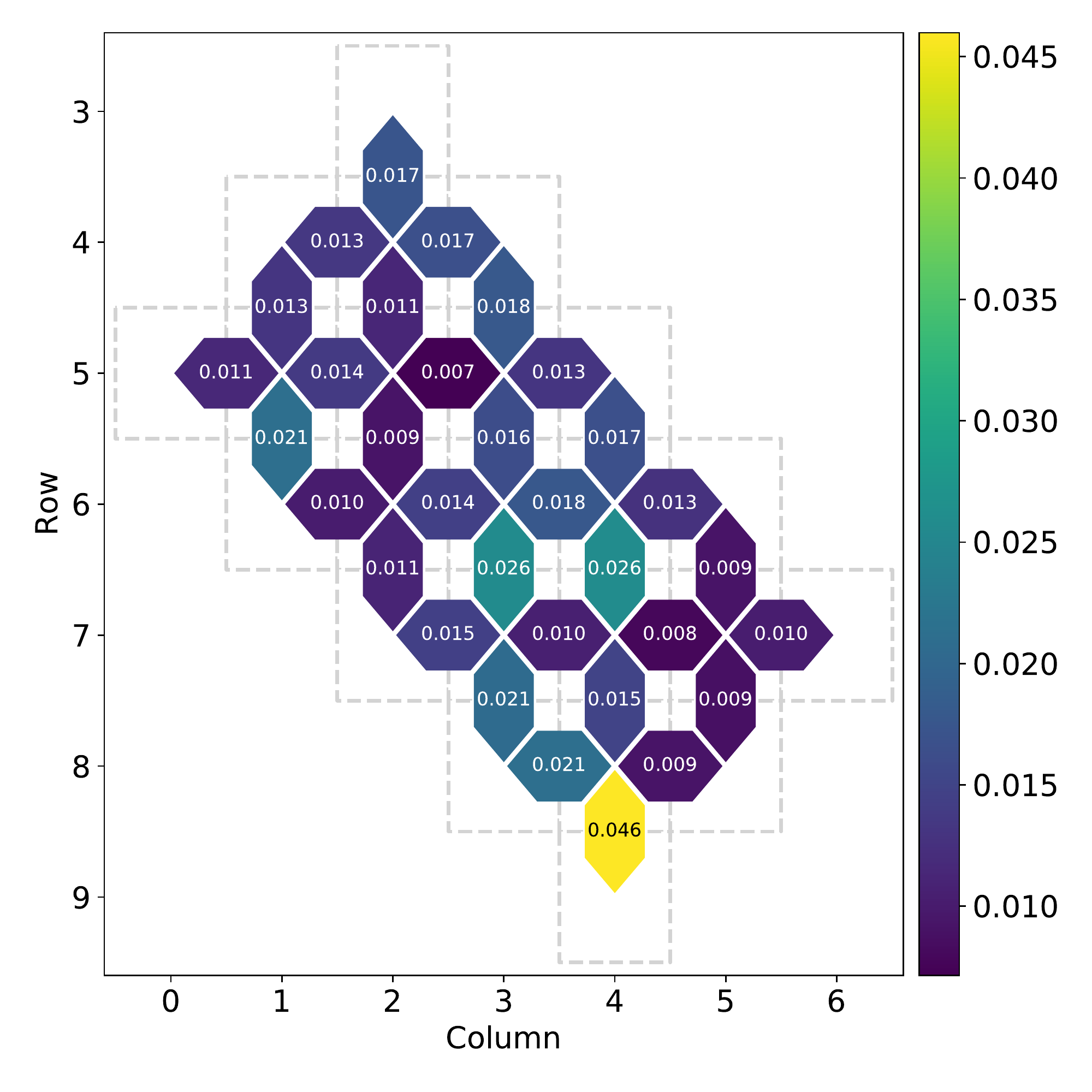}
\caption{\label{fig:two_qubit_error_rain} \textbf{Rainbow parallel $\sqrt{\text{iSWAP}}^{\dagger}$ Pauli error rates.} Same data as Fig.~\ref{fig:two_qubit_error}, but for the 23-qubit Rainbow quantum processor.
}
\end{figure}

The Rainbow quantum processing unit is a 23-qubit superconducting processor with typical performance characteristics similar to Weber. While the results in the main text were generated on the Weber processor, we used Rainbow in order to assess the effect of varying initial conditions in the product state. These varying initial conditions were assessed in a 17-qubit chain, the largest 1D chain embeddable on Rainbow. This size of the chain is large enough to not be susceptible to the finite-size effects encountered in smaller chains.  At the same time it is not so large that low retained count fractions cause problems with measurement statistics and thus create large error bars. Figs.~\ref{fig:readout_error_rain}, \ref{fig:t1_time_rain}, and \ref{fig:two_qubit_error_rain} show the three most relevant performance characteristics for choosing high-quality chains, parallel $e_{r1}$, $\tilde{T}_1$, and parallel $e_2$, respectively. However, the ability to avoid noisy qubits on Rainbow was severely constrained by the large size of the chain relative to the size of the processor. For reference, we used the embedding (5, 0) $\rightarrow$ (5, 1) $\rightarrow$ (4, 1) $\rightarrow$ (4, 2) $\rightarrow$ (4, 3) $\rightarrow$ (5, 3) $\rightarrow$ (5, 2) $\rightarrow$ (6, 2) $\rightarrow$ (7, 2) $\rightarrow$ (7, 3) $\rightarrow$ (6, 3) $\rightarrow$ (6, 4) $\rightarrow$ (6, 5) $\rightarrow$ (7, 5) $\rightarrow$ (7, 4) $\rightarrow$ (8, 4) $\rightarrow$ (8, 5).

\section{Circuit Compilation, Calibration, and Optimization}

In order to simulate the $T_6$ QCA with high-fidelity, we compile, calibrate, and optimize circuits in the following manner. The Cirq open source framework was used for the workflow \cite{cirq_developers_2021_5182845}.

\begin{enumerate}
    \item Each $CH(q_i, q_j)$ gate between control qubit $q_i$ and target qubit $q_j$ in the local $T_6$ update unitary is compiled into
    \begin{equation}
    CH(q_i, q_j) = Y^{1/4}(q_j) CZ(q_i, q_j) Y^{-1/4}(q_j),
    \end{equation}
    where up to a global phase $Y^{t}=R_Y(\pi t)$.
    
    \item Each $CZ$ gate is further decomposed into two bare $\sqrt{\text{iSWAP}}^{\dagger}$ gates and single-qubit rotations. A controlled-phase gate between control qubit $q_i$ and target qubit $q_j$ can be decomposed as
    \begin{align}\label{eq:cphase_decomp}
        \begin{split}
            \mathrm{CPHASE}(\phi)_{ij}&=e^{i(2\varphi-\phi)/4}\\
            \times &R_{Z_i}(\pi-\phi/2) \otimes R_{Z_j}(-\phi/2)\\
            \times &R_{X_i}(-\xi_i) \otimes R_{X_j}(-\xi_j) \\
            \times &K(\theta)_{ij}\mathrm{CPHASE}(\varphi)_{ij}\\
            \times &R_{Z_i}(\pi+\varphi/2)\otimes R_{Z_j}(\varphi/2)\\
            \times &R_{X_i}(-2\alpha) \otimes 1_j\\
            \times &K(\theta)_{ij}\mathrm{CPHASE}(\varphi)_{ij}\\
            \times &R_{Z_i}(\phi/2) \otimes R_{Z_j}(\phi/2)\\
            \times &R_{X_i}(\xi_i)\otimes R_{X_j}(\xi_j),
        \end{split}
    \end{align}
    which is equivalent to the decomposition presented in Ref. \cite{arute2020observation}. Here, $K(\theta)$ is a continuously-parameterized fractional-iSWAP gate such that $\sqrt{\text{iSWAP}}^{\dagger}=K(\pi/4)$. In this step, we take $\varphi=0$ and account for the parasitic $\varphi \approx \pi/23$ using Floquet calibration in a later step. The other decomposition parameters are given by
    \begin{equation}
        \sin(\alpha) = \sqrt{\frac{\sin^2(\phi/4)-\sin^2(\varphi/2)}{\sin^2(\theta)-\sin^2(\varphi/2)}},
    \end{equation}
    \begin{align}
    \begin{split}
        \xi_i &= \tan^{-1}\Bigg(\frac{\tan(\alpha)\cos(\theta)}{\cos(\varphi/2)}\Bigg)\\
        &+\frac{\pi}{2}\Big(1-\text{sgn}\Big(\cos(\varphi/2)\Big)\Big),
    \end{split}
    \end{align}
    and
    \begin{align}
    \begin{split}
        \xi_j &= \tan^{-1}\Bigg(\frac{\tan(\alpha)\sin(\theta)}{\sin(\varphi/2)}\Bigg)\\
        &+\frac{\pi}{2}\Big(1-\text{sgn}\Big(\sin(\varphi/2)\Big)\Big).
    \end{split}
    \end{align}
    To decompose the $CZ$ gate, we simply set $\phi=\pi$.
    
    \item Strings of consecutive single-qubit gates are merged together into $\text{PhXZ}(a, x, z) \equiv Z^z Z^a X^x Z^{-a}$ gates using the cirq.google \textit{optimized\_for\_sycamore} utility. This compresses into alternating moments of parallel single-qubit gates followed by parallel two-qubit gates.
    
    \item Spin echo insertion: Wherever a qubit is idle for more than one single-qubit gate layer, pairs of Pauli operators are inserted to decrease idle qubit crosstalk. For instance, $XX=1$.
    
    \item Floquet characterization: Our targeted native two-qubit gate on Weber is the $\sqrt{\text{iSWAP}}^{\dagger}=K(\pi/4)$ gate. However, due to calibration drift and cross-talk errors, the effective gate implemented has the form (see Supplementary Information of Refs.~\cite{arute2020observation, neill2021accurately})
    \begin{equation}
        \begin{split}
            \quad \quad & K(\theta,\zeta, \chi,\gamma,\varphi) = \\
            & \begin{bmatrix}
                    1 & 0 & 0 & 0\\
                    0 & e^{-i\gamma-i\zeta}\cos(\theta) & -i e^{-i\gamma +i\chi}\sin(\theta) & 0\\
                    0 & -i e^{-i\chi - i\gamma}\sin(\theta) & e^{-i\gamma+i\zeta}\cos(\theta) & 0 \\
                    0 & 0 & 0 & e^{-2i\gamma - i\varphi}
             \end{bmatrix}.
        \end{split}
    \end{equation}
    In the absence of calibration drift and cross-talk errors, $\theta=\pi/4$ and $\zeta=\chi=\gamma=\varphi=0$, and one recovers the target gate as $\sqrt{\text{iSWAP}}^{\dagger}=K(\pi/4, 0, 0, 0, 0)$. When such drift and errors are present, the goal of Floquet characterization is to determine the values of $\theta,\zeta, \chi,\gamma,$ and $\varphi$ with a view towards subsequently correcting for some or all of them. This is done by repeating circuit moments with two-qubit gates many times with interleaved, parameterized z-rotations as probes, in order to amplify the calibration drift and control errors embodied in $\theta,\zeta, \chi,\gamma,$ and $\varphi$. All five angles except for $\chi$ are able to be characterized, since $\chi$ corresponds to a complex hopping phase and requires closed loops in order to characterize. For more details, see Refs. \cite{arute2020observation, neill2021accurately}.
    
    \item After Floquet characterization, the first angle we correct for is the parasitic cphase angle, $\varphi$. We do this by reconstructing our circuits and re-decomposing our $CZ$ gates using Eq.~\ref{eq:cphase_decomp} where now instead of $\varphi=0$ we use the value determined by Floquet characterization on each moment, which is typically in the vicinity of $\varphi\approx \pi/23$.
    
    \item Next, we re-merge single-qubit gates according to the procedure in Step 3.
    
    \item Re-insert spin echoes according to the logic of Step 4.
    
    \item Floquet calibration: In addition to $\varphi$, the angles $\gamma$ and $\zeta$ can be corrected for by inserting z-rotations on either side of an imperfect two-qubit gate. The relationship between the general excitation-number-conserving gate and our target gate (with parasitic cphase) is
    \begin{equation}\label{eq:angle_correction}
        \begin{split}
        \quad \quad &K(\theta, 0, 0, 0, \varphi) = \\
        &R_Z(-\beta,\beta)R_Z(\gamma, \gamma)K(\theta,\zeta,\chi,\gamma,\varphi)R_Z(-\alpha, \alpha),
        \end{split}
    \end{equation}
    where $\alpha=(\zeta+\chi)/2$, $\beta=(\zeta-\chi)/2$, and we use the following shorthand for pairs of two single-qubit z-rotations, $R_Z(z_i,z_j)=e^{i(z_i+z_j)/2}R_{Z_i}(z_i)\otimes R_{Z_j}(z_j)$. Since we cannot characterize $\chi$, we set $\chi=0$. The resulting simplification in Eq.~\ref{eq:angle_correction} will correct for $\gamma$ and $\zeta$. As such, after the insertion of the z-rotations in Eq.~\ref{eq:angle_correction}, each $CZ$ gate in the original QCA circuit is compiled correctly as in Eq.~\ref{eq:cphase_decomp} with the effective two-qubit native gate $\sim K(\pi/4,0,0,0,\pi/23) = K(\pi/4)\mathrm{CPHASE}(\pi/23)$. The angles $\theta\sim\pi/4$ and $\chi\sim 0$ remain uncorrected. The inserted, corrective z-rotations are merged with the execution of two-qubit gates at the hardware level and do not need to be re-merged into $\text{PhXZ}$ gates.
    
    \item Send circuits to the quantum processor, Weber or Rainbow, in temporal batches, averaging over qubit arrangements.
\end{enumerate}

\section{Post-Selection}

\begin{figure}
\includegraphics[width=.9\linewidth]{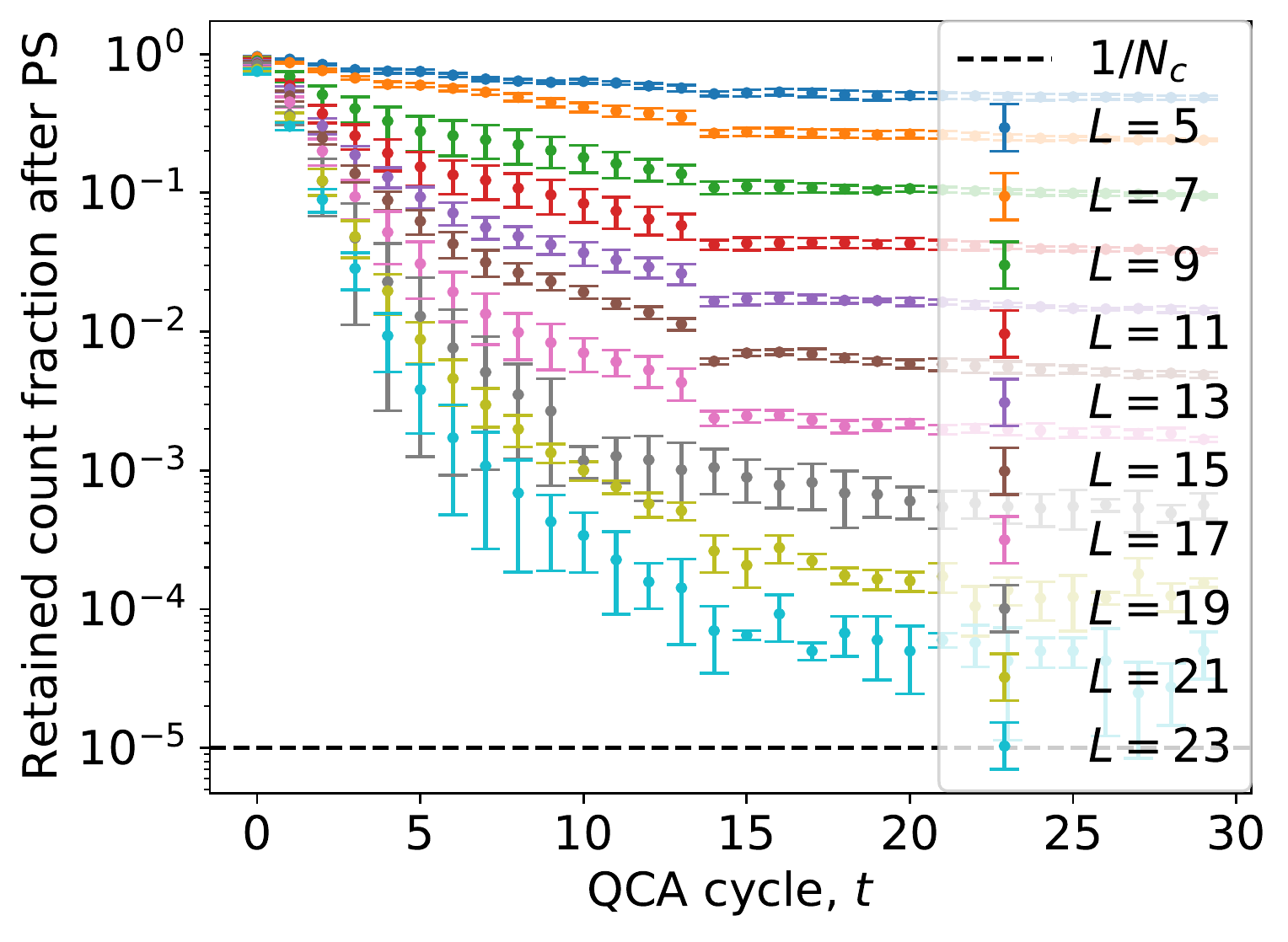}
\caption{\label{fig:ps_overhead} \textbf{Fraction of $z$-basis measurements retained after post-selection.} 
}
\end{figure}

Once measurement statistics are gathered by measuring $N_c$ times in the $z$-basis, post-selection is used to discard measurements, i.e. bit strings, that result from error and thus reside outside the invariant-protected sector of Hilbert space. We emphasize that with a higher fidelity, post-selection would be unnecessary for the establishment of coherence windows; generally however, in the NISQ era post-selection is an important and useful tool for quantum computing. The dynamical invariant that defines the protected sector of Hilbert space is the operator
\begin{equation}\label{eq:invariant}
    \begin{split}
        \mathcal{O} \equiv \sum_{i=0}^L \mathcal{O}_i,\\
        \mathcal{O}_i \equiv Z_i Z_{i+1}
    \end{split}
\end{equation}
where the sites $i=0$ and $i=L+1$ refer to padding boundary qubits fixed to the $|0\rangle$ state. We first give a proof that $\mathcal{O}$ is a dynamical invariant of the $T_6$ rule. Consider the commutator between Eq.~\ref{eq:qca_unitary} and Eq.~\ref{eq:invariant}
\begin{equation}\label{eq:new_full_comm}
    \begin{split}
        \Big[ \mathcal{O}, U(T_R;t,t+1) \Big] = \sum_{i=0}^L \Big[\mathcal{O}_i,\prod_o U_o \prod_e U_e \Big]\\
        = \sum_{i=0}^L \Big( \Big[\mathcal{O}_i,\prod_o U_o \Big]\prod_e U_e + \prod_o U_o \Big[\mathcal{O}_i,\prod_e U_e \Big] \Big).
    \end{split}
\end{equation}
Using standard commutator identities, the two commutators in the second line of Eq.~\ref{eq:new_full_comm} can be expressed as
\begin{equation}\label{eq:half_comm}
    \begin{split}
        \Big[\mathcal{O}_i, \prod_j U_j \Big]=\sum_j \prod_{l'>j} U_{l'}\Big[\mathcal{O}_i,U_j\Big]\prod_{l<j}U_l\\
        =\sum_j \prod_{l'>j} U_{l'}\Big( Z_i\Big[Z_{i+1},U_j\Big]+\Big[Z_{i},U_j\Big]Z_{i+1}\Big)\prod_{l<j}U_l,
    \end{split}
\end{equation}
where $j$ can run over either the even or odd index set. Next, note that since any projection operator commutes with the Pauli $Z$ operator,
\begin{equation}\label{eq:term_comm}
    \begin{split}
        \Big[ Z_{k} , U_j \Big] = P_{j-1}^{(m)}\Big[ Z_j, V^{c_{mn}}_j \Big] \delta_{k,j} P_{j+1}^{(n)}.
    \end{split}
\end{equation}
Substituting Eqs.~\ref{eq:term_comm} and~\ref{eq:half_comm} into Eq.~\ref{eq:new_full_comm} and noting that $Z_k P_k^{(m)} = P_k^{(m)} Z_k = (-1)^m P_k^{(m)}$ we find
\begin{equation}
    \begin{split}
        \Big[ \mathcal{O}, U(T_R;t,t+1) \Big] =\\ \sum_o \prod_{l'>o} U_{l'}\sum_{m,n}P_{o-1}^{(m)}\Big[Z_o,V_o^{c_{mn}}\Big]P_{o+1}^{(n)}\Big( (-1)^m \\ + (-1)^n \Big)\prod_{l<o}U_l \prod_e U_e\\
        + \prod_o U_o \sum_e \prod_{l'>e} U_{l'}\sum_{m,n}P_{e-1}^{(m)}\Big[Z_e,V_e^{c_{mn}}\Big]P_{e+1}^{(n)}\Big( (-1)^m \\ + (-1)^n \Big)\prod_{l<e}U_l
    \end{split}
\end{equation}
We now specify to the rule $T_6$. There are four combinations of $m$ and $n$ to consider. In the instance where $m=n=0$ or $m=n=1$, the commutator $[Z_j,V_j^{c_{m=n}=0}]=0$. Meanwhile, when $m\neq n$, $(-1)^m + (-1)^n = 0$. Therefore, each term indexed by $(m,n)$ vanishes and
\begin{equation}
    \Big[ \mathcal{O}, U(T_6;t,t+1) \Big]=0,
\end{equation}
proving that $\mathcal{O}$ is a dynamical invariant of rule $T_6$ for \textit{any} unitary $V_j$.

In addition, $\mathcal{O}$ has eigenstates diagonal in the $z$-basis. Hence, any $z$-basis measurement whose eigenvalue under $\mathcal{O}$ is different than the $\mathcal{O}$-eigenvalue of the QCA's initialized state must have resulted from an error in the computation. In order to see this, and that Eq.~\ref{eq:invariant} is related to domain wall conservation, consider the action of $\mathcal{O}$ on a five-qubit chain, initialized with a single, central bit flip, and padded by fixed boundary $|0\rangle$s
\begin{equation}
    \mathcal{O}\,|0\rangle \!\otimes\! |00100\rangle \!\otimes\! |0\rangle = 2\, |0\rangle \!\otimes\! |00100\rangle \!\otimes\! |0\rangle.
\end{equation}
Now consider the action of $\mathcal{O}$ on another state with the same number of domain walls
\begin{equation}\label{eq:good_state}
    \mathcal{O}\,|0\rangle \!\otimes\! |01110\rangle \!\otimes\! |0\rangle = 2\, |0\rangle \!\otimes\! |01110\rangle \!\otimes\! |0\rangle.
\end{equation}
If however a bit flip were to occur on the central qubit in the chain, the action of $\mathcal{O}$ would be
\begin{equation}
    \mathcal{O}\,|0\rangle \!\otimes\! |01010\rangle \!\otimes\! |0\rangle = -2\, |0\rangle \!\otimes\! |01010\rangle \!\otimes\! |0\rangle.
\end{equation}
Hence, the calculated eigenvalue of $-2$, rather than $+2$, indicates that we should discard this measurement from our statistics when calculating observables.

Note that $\mathcal{O}$ does not protect against all longitudinal relaxation errors. For example, if the state in Eq.~\ref{eq:good_state} relaxes on the qubit just left or right of center, $\mathcal{O}$ still has eigenvalue $+2$. In addition, if two photons are lost, say in both the central qubit and in the one just to the right or left, $\mathcal{O}$ will also have eigenvalue $+2$. In instances such as these, superposition amplitudes will renormalize in such a way as to degrade the fidelity of the computation while still keeping the processor's state superposition in the protected subspace.

Fig.~\ref{fig:ps_overhead} shows the fraction of initial processor counts, $N_c$, that are retained after applying post-selection to our experimental measurements from Weber as a function of QCA cycle depth for all system sizes considered, $L=5, 7, 9, \ldots, 23$. Unsurprisingly, from $t=0$ to $t=14$, each set of data points with constant $L$ decays exponentially as a function of QCA cycle as the post-selection procedure has to discard exponentially many measurements in order to compensate for $\tilde{T}_1$ decoherence. At $t=15$, there is an evident step change in the decay rate of the retained count fractions. The nominal gate execution times for $\text{PhXZ}$ and $\sqrt{\text{iSWAP}}^{\dagger}$ on Weber are 25 ns and 32 ns, respectively. Since each QCA cycle has eight alternating layers of each, at $t=15$, the nominal circuit execution time (excluding readout) is about $6.84 \: \mu \text{s}$. While median $\tilde{T}_1$ times are about $15 \: \mu \text{s}$, low end (10th percentile) $\tilde{T}_1$ times are about $11 \: \mu \text{s}$ on Weber. Hence, at $6.84 \: \mu \text{s}$, the probability of a 10th percentile qubit erroneously relaxing from the $|1\rangle$ state is about $47 \: \%$, making it difficult for post-selection to compensate for multi-photon loss processes or delocalizing errors. QCA cycle $t=15$ is also the point referenced in the main text (e.g., main text Fig. 2b) where the error bars on population dynamics begin to grow significantly.

In addition, the retained count fraction also decays roughly exponentially at fixed $t$ as a function of $L$. The number of gate layers, and thus the circuit execution time, is fixed as a function of system size for fixed QCA cycle depth with respect to $\tilde{T}_1$. Since gate \textit{volume} however grows linearly in $L$, it is likely that this decay in the retained count fraction as a function of system size is a result of accumulating gate error, as is well-established within the digital error model \cite{arute2019quantum}.

\begin{figure}
\includegraphics[width=\linewidth]{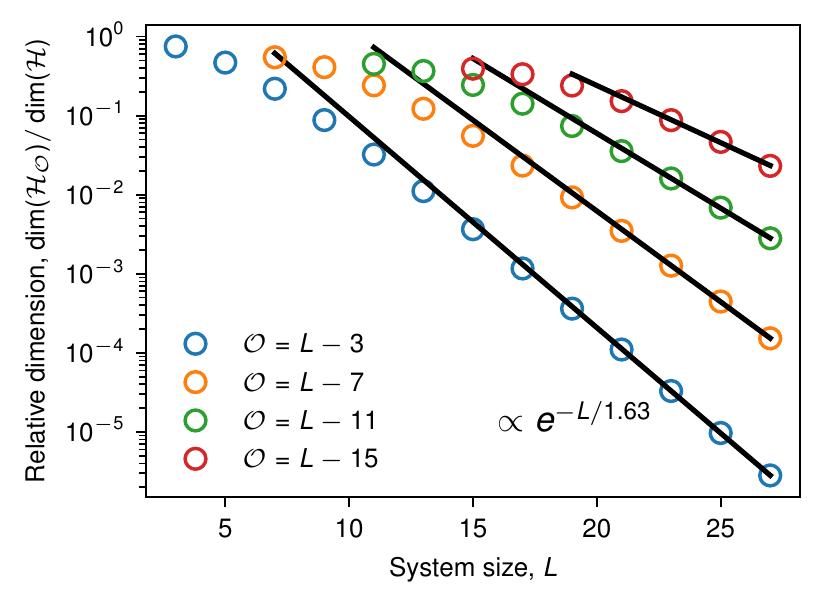}
\caption{\label{fig:hilbert_dim} \textbf{Relative Hilbert space dimension} of sectors with constant $\mathcal{O} = L - m$ for $m \in \{3, 7, 11, 15\}$. These sectors correspond to initial conditions with 1, 2, 3, and 4, $|1\rangle$'s that are well separated by $|0\rangle$'s, respectively. Normalization is provided by the total Hilbert space size $2^L$. Black lines are exponential fits to the last four data points.
}
\end{figure}
Since $T_6$ conserves $\mathcal{O}$, the system evolves in a Hilbert space of reduced dimension $\mathcal{H}_{\mathcal{O}}$.  In Fig.~\ref{fig:hilbert_dim} we plot the relative dimension $\dim(\mathcal{H}_{\mathcal{O}}) / \dim(\mathcal{H}$) for initial conditions consisting of a few $|1 \rangle$'s separated by $|0 \rangle$'s. For the case studied in the main text, a single $|1\rangle$, the relative Hilbert space dimension decays as $\propto e^{-L/1.63}$. Thus, since $\dim(\mathcal{H})=2^L$, we have $\dim(\mathcal{H}_{L-3}) \propto 1.08^L$.

\section{Mutual Information Measures}

\begin{figure}
\includegraphics[width=1.\linewidth]{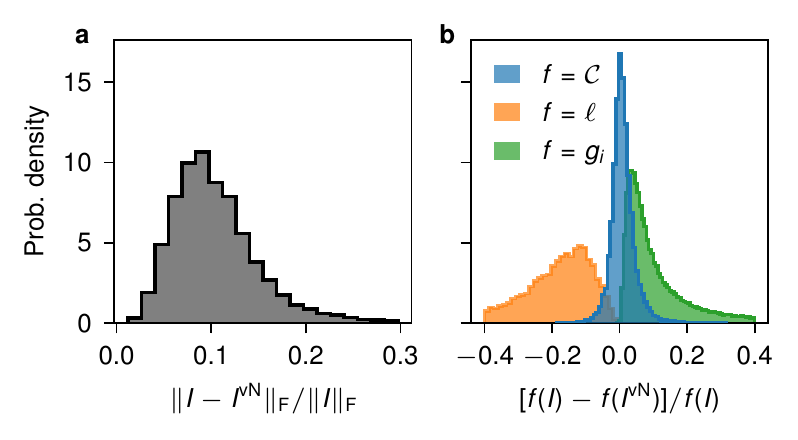}
\caption{\label{fig:shannon_v_vN} \textbf{Comparing Shannon and von Neumann mutual information.} Distribution of {\bf a}, relative Frobenius distances and {\bf b}, relative network measure differences over 10,000 QCA cycles.
}
\end{figure}

Consider a set of $z$-basis measurements $\{|z\rangle \: \: \text{s.t.} \in \{0, 1\}^{\otimes N}\}$ from an $N$-qubit (subset of a) quantum processor and associated probabilities, $\{P_z\}$. For any two qubits, $q_i$ and $q_j$, one can construct a $z$-basis joint and two marginal probability distributions via
\begin{align}
    p(z_i,z_j) &= \sum_{z'} P_{z'} \delta_{z_i,z_i'} \delta_{z_j,z_j'}\\
    p(z_{i(j)}) &= \sum_{z_{j(i)}} p(z_i,z_j).
\end{align}
For two classical variables (e.g., bits) $z_i$ and $z_j$ with joint probability distribution $p(z_i,z_j)$ and marginal probability distributions, $p(z_i)$ and $p(z_j)$, the Shannon mutual information is defined in terms of the Shannon entropy of a distribution, ${\rm H}(p)$, as (see e.g., Ref.~\cite{nielsen2002quantum})
\begin{equation}
    I_{ij} \equiv {\rm H}[p(z_i)] + {\rm H}[p(z_j)] -{\rm H}[p(z_i,z_j)]
\end{equation}
and can be written as
\begin{equation}
        I_{ij} \equiv \sum_{z_i=0}^1 \sum_{z_j=0}^1 p(z_i, z_j) \log_2 \frac{p(z_i, z_j)}{p(z_i)p(z_j)}.
\end{equation}
It measures, as a relative entropy, correlations in the variables $z_i$ and $z_j$.
That is, $I_{ij}\geq 0$ and $I_{ij}=0$ if and only if $p(z_i,z_j)=p(z_i)p(z_j)$.

The quantum mechanical generalization of the classical probability distributions is the reduced density matrix (RDM). For a Hilbert space composed of two subspaces, $\mathcal{H}=\mathcal{H}_A \otimes \mathcal{H}_B$ and a density matrix $\rho \in \mathcal{H}$, the RDM of subspace $\mathcal{H}_A$ is defined as
\begin{equation}
    \rho_A \equiv \text{Tr}_B \rho,
\end{equation}
where the the trace is taken over the degrees of freedom of subsystem $\mathcal{H}_B$. RDMs can also be constructed from few-body quantum observables. For instance, the single-qubit RDM for qubit $q_i$ can be constructed as
\begin{equation}
    \rho_i = \frac{1}{2}\sum_{\mu=0}^3 \langle \sigma_i^{\mu} \rangle \sigma_i^{\mu},
\end{equation}
where $\sigma_i^0=1_i$, $\sigma_i^1=X_i$, $\sigma_i^2=Y_i$, and $\sigma_i^3=Z_i$ are elements of the Pauli algebra, and expectation values are calculated from quantum processor measurements in the appropriate basis \cite{gamel2016entangled}. Similarly, a $k$-qubit RDM can be calculated as
\begin{equation}
    \rho_{i_1\ldots i_k}=\frac{1}{2^k}\sum_{\mu_1,\ldots,\mu_k=0}^3 \langle \sigma_{i_1}^{\mu_1}\otimes \ldots \otimes \sigma_{i_k}^{\mu_k} \rangle \sigma_{i_1}^{\mu_1}\otimes \ldots \otimes \sigma_{i_k}^{\mu_k}.
\end{equation}
Expectation values of long Pauli strings can be efficiently calculated via schema such as quantum overlapping tomography \cite{cotler2020quantum}. With an RDM in hand, the von Neumann entropy is defined as
\begin{equation}
    S(\rho) \equiv - \text{Tr} \rho \log_2 \rho
\end{equation}
and the von Neumann mutual information between two qubits follows similarly
\begin{equation}
    I_{ij}^{\rm vN} \equiv S(\rho_i) + S(\rho_j) - S(\rho_{ij}).
\end{equation}
Calculation of $I_{ij}^{\rm vN}$ therefore requires measurements of observables, such as $\langle X_i Z_j \rangle$, in bases other than the $z$-basis. However, the dynamical invariant introduced in the main text, $\mathcal{O}=\sum_{i=0}^L Z_i Z_{i+1}$, can only be used to post-select for errors when measurement is performed in the $z$-basis. As such, we choose to use Shannon mutual information for the calculation of complex network measures. 

As validation for this choice, we numerically emulate $L=19$ qubits initialized with a single-bit flip and evolving under $T_6$ for 10,000 cycles. At each cycle we calculate the relative Frobenius distance $\Vert I - I^{\rm vN} \Vert_{\rm F} / \Vert I \Vert_{\rm F}$, where the Frobenius norm of a matrix $M$ with elements $M_{ij}$ is defined as
\begin{equation}
    \Vert M \Vert_{\rm F} = \sqrt{\sum_{i,j=0}^{L-1} \vert M_{ij}\vert^2}.
\end{equation}
Fig.~\ref{fig:shannon_v_vN}a shows the distribution of the relative Frobenius distance between the Shannon and Von Neumann mutual information across 10,000 cycles. The median relative distance is $9.7\%$. We also examine relative differences of network measures (defined in detail in the next section) in Fig.~\ref{fig:shannon_v_vN}b. Clustering is the most similar between Shannon and von Neumann-based mutual information (blue histogram, median relative difference $0.7\%$). Path length (gold histogram) exhibits a median relative difference $-18\%$) while that of node strength (green histogram) is $8\%$. We conclude that the Shannon mutual information is a reliable proxy the the von Neumann entropy at the level of, at worst, a few percent. Moreover, since we use only the Shannon mutual information in the main text, our conclusions regarding QCA network complexity, as compared to random networks and post-selected incoherent uniform random states, stand irrespective of the mutual information definition employed.

\section{Complex Network Measures}

\subsection{Clustering Coefficient}

In order to establish the small-world character of the $T_6$ QCA's mutual information network, we focus on three canonical complex network measures: clustering, path length, and node strength distribution. Intuitively, one can view the mutual information matrix $I_{ij}$, as an adjacency matrix in analogy with that of a transportation network. For example, the economic activities of two cities are likely to be more tightly intertwined if high-speed rail (large adjacency weight) connects them rather than a rutted dirt track (low adjacency weight). Similarly, the dynamics of two qubits will be more strongly correlated if they share more mutual information.

Continuing with this analogy, the clustering coefficient gauges how locally traversable a network is. Suppose Alice is attempting to reach City C from City A, but there only exist rail lines from City A to City B and from City B to City C, but no lines exist, or perhaps just a dirt track exists, between City A and City C. Locally, this is an inefficient network to traverse since Alice either has to travel through City B else has to take the dirt track to get between Cities A and C. A highly traversable network at the local scale also implies high network transitivity; that is, if Cities A and B are well-connected and Cities B and C are well-connected, then so should be cities A and C. The \textit{global clustering coefficient} assesses this local transitivity in a manner that averages over the entire network. At a particular cycle depth, $t$, it is defined as
\begin{equation}\label{eq:clustering}
    \mathcal{C} \equiv \frac{\text{Tr}[I^3]}{\sum_{i\neq j=1}^L [I^2]_{ij}}.
\end{equation}
The numerator of Eq.~\ref{eq:clustering} counts the weighted number of closed triangles (node triplets) in the network, while the denominator is the weighted number of length-2 paths in the network, that is, the weighted number of potentially closed triangles. 
In the main text, we show that the Goldilocks rule $T_6$ exhibits sizeable clustering, much larger than the post-selected uniform random state.   The clustering grows with system size within the Weber processor coherence window, indicating experimental formation of a series of locally traversable networks.

\subsection{Average Shortest Path Length}

A network's \textit{average shortest path length} measures the extent to which it is globally traversable. In the transportation network analogy, a network of cities (e.g., Atlanta, Nashville, Indianapolis, Detroit, Pittsburgh, Baltimore, Raleigh, and Charlotte) laid out in a one-dimensional ring topology but only connected to their nearest neighbors (Atlanta and Indianapolis for Nashville) and next-nearest neighbors (Charlotte and Detroit for Nashville) via high-speed rail lines would have very high local traversability, i.e., transitivity. However, it would take a relatively long time to traverse between antipodes of the ring (e.g., Nashville to Baltimore) and thus those cities' economic activities might be less correlated. From the standpoint of network weights, a large mutual information between two nodes should contribute to shortening path length, such as a rail line that could carry more passengers might, while small mutual information should increase the path length by virtue of its low traversability, like the dirt track. We therefore define the network-averaged weighted path length for a particular cycle depth as
\begin{equation}
\begin{aligned}
    \ell \equiv \frac{1}{L(L-1)} &\sum_{i\neq j=1}^L d_{ij},\\
    d_{ij} = \min_{p_{ij} \in \mathcal{P}_{ij}} &\sum_{\langle k, l \rangle \in p_{ij}} 1/I_{kl},
\end{aligned}
\end{equation}
where $d_{ij}$ is the minimum distance between nodes (i.e., qubits) $q_i$ and $q_j$, the sum in the first line runs over all pairs of qubits, and $L(L-1)$ is the number of pairs in the network. The sum in the second line runs over all edges ($\langle k,l \rangle$) in a particular path ($p_{ij}$) between nodes $q_i$ and $q_j$ and the minimum is taken over all possible paths ($\mathcal{P}_{ij}=\{p_{ij}\}$) between the two nodes. As discussed, we use $1/I_{kl}$ as weights in the summand under the intuition that edges with large mutual information are traversed easily while small mutual information values should be detrimental to short path length, consistent with \cite{muldoon2016small}.

\subsection{Node-Strength Distribution}

The last complex network measure we consider is the \textit{node-strength distribution}. In the transportation network analogy, an efficiently traversable network will have many nodes like Grand Central Station (termed \textit{hubs}) that have many, strong connections to other nodes as well as some nodes that have few or weak connections, like Golden, Colorado, whose servicing light rail line terminates three miles short of the center of town. Mathematically, this translates into a broad, flat node-strength distribution. We calculate a size-invariant (un-normalized) node-strength distribution,
\begin{equation}\label{eq:node_strength}
    P[g_i/(L-1)] = \text{hist}\Bigg[ \frac{1}{L-1} \sum_j I_{ij} \Bigg]
\end{equation}
Practically, Eq.~\ref{eq:node_strength} is evaluated by aggregating all size-scaled values of $g_i/(L-1)=\sum_{j}I_{ij}/(L-1)$ across all system sizes, $L\in\{5, 7, \ldots, 23\}$ into a set which is then histogrammed. The resulting distribution is increasingly biased towards low node strength as the network approaches randomness. Small-world networks have broad, flat distributions, while distributions that are more sharply-peaked around smaller ranges of values indicate increasing non-random regularity in the network.

\section{Establishing the Coherence Window}

\begin{figure}
\includegraphics[width=1.\linewidth]{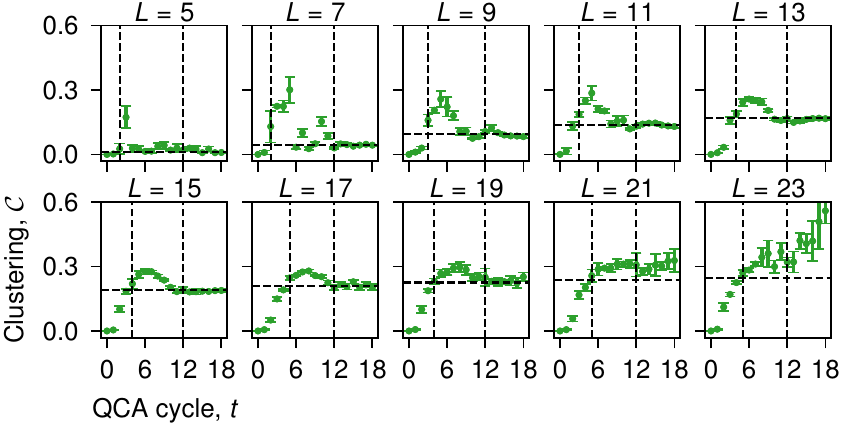}
\caption{\label{fig:coherence_window} \textbf{Establishing the coherence window.} Clustering coefficient calculated from the post-selected data generated from the Weber processor (green data points), as a function of QCA cycle and for all system sizes simulated (one in each panel), $L\in \{5, 7, \ldots, 23\}$. Error bars: one standard deviation from the mean on four different qubit chains. Horizontal dashed lines: clustering coefficient for post-selected incoherent uniform randomness, based on the QCA initial condition of a single $|1\rangle$ initialized in a chain of $L$ qubits. Vertical dashed lines: beginning and end of the coherence window in each subplot.
}
\end{figure}

As discussed in the main text, our Goldilocks QCA circuits, when aided by post-selection, generate mutual information complex network observables that exhibit structure in excess of post-selected incoherent uniform randomness. Fig.~\ref{fig:coherence_window} shows the clustering coefficient calculated from the post-selected data generated from the Weber processor (green data points), as a function of QCA cycle and for all system sizes simulated, $L\in \{5, 7, \ldots, 23\}$. Over the first few cycles, the clustering coefficient climbs from zero until it crosses the horizontal dashed line in each panel, which is the value of the clustering coefficient for the incoherent uniform random state subject to the same post-selection procedure as the QCA data from Weber. The left-most vertical dashed line demarcates this point, which is the QCA cycle at which the post-selected experimental clustering becomes larger than post-selected randomness. In all cases, the post-selected experimental clustering continues to climb for a few cycles until decoherence mechanisms begin to degrade the data in excess of what post-selection can mitigate for, typically at $t\approx 8$. After a few more cycles, the post-selected experimental clustering drops to the clustering value of post-selected incoherent uniform randomness and remains there for the rest of the simulation. The right-most vertical dashed line demarcates this point, which occurs at $t\approx 12$ roughly independent of system size. For each system size, the region between the two vertical dashed lines is where the post-selected experimental clustering is in excess of post-selected incoherent uniform randomness and is therefore the \textit{coherence window} over which we calculate the cycle-averaged complex network observables. We remark that after the end of the coherence window, $t\approx 12$, for $L=21$ and $L=23$ the post-selected experimental clustering fails to return to the value associated with post-selected incoherent uniform randomness. We do not regard this as an extension of the coherence window. Rather, inspection of Fig.~\ref{fig:ps_overhead} shows that after $t\approx 12$, the 21-qubit chains are retaining fewer than roughly 100 of the 100,000 initial measurements while the 23-qubit chains are retaining fewer than about 10 of the initial 100,000 measurements. As such, error bars become significant and calculation of observables after $t\approx 12$ in each case becomes unreliable.

\section{Effect of Higher Product State Filling}

\begin{figure*}
\includegraphics[width=.95\linewidth]{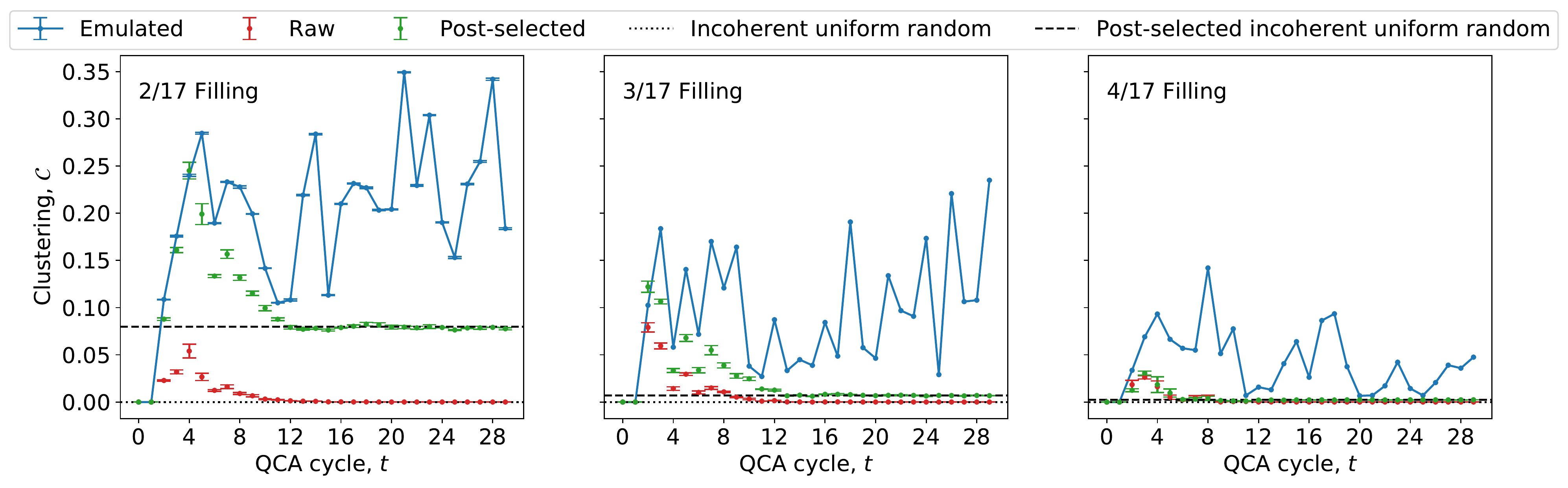}
\caption{\label{fig:clusterings_ic} \textbf{Effect of higher product state filling.} Clustering coefficient as a function of QCA cycle for three different numbers of isolated $|1\rangle$s in the initial product state, 3, 4, and 5, for $L=17$ on the Rainbow processor (blue curves: numerical emulation; red points: Rainbow data without post-selection; green points: with post-selection). Error bars are one standard deviation in $\mathcal{C}$ over four different qubit chains. Black dashed lines represent the clustering of the incoherent uniform random state, subject to post-selection on the appropriate eigenvalues of $\mathcal{O}$ corresponding to the three different initial conditions: 10 ($L-7$), 6 ($L-11$), and 2 ($L-15$).
}
\end{figure*}

Throughout the main text, we have focused on a particular initial condition for our Goldilocks QCA circuits, namely $|0\ldots 010 \ldots 0\rangle$, a single central bit flip in a chain of $L$ qubits. This was done so as to be able to perform finite size scaling analyses of complex network measures in a sensible and well-controlled manner, as well as the close analogy to a standard initial condition for elementary classical cellular automata. Moreover, it was previously shown that the ability of Goldilocks QCA to generate physical complexity is largely insensitive to initial conditions \cite{hillberry2021entangled}. However, it is an important question as to whether our experimental protocol, including post-selection, for effectively generating coherent small-world mutual information networks generalizes beyond the particular instance of the single bit-flip initialization.

Fig.~\ref{fig:clusterings_ic} shows the effect of increasing the number of isolated bit flips in the initial state on clustering dynamics of a 17-qubit chain simulated on Google's 23-qubit Rainbow processor. In particular, the left panel corresponds to the initial condition $|00000100000100000\rangle$, the middle panel corresponds to the initial condition $|00010000100001000\rangle$ and the right panel corresponds to the initial condition $|00100010001000100\rangle$, i.e., equally spaced $|1\rangle$'s. We observe the following.  First, the long-time average clustering values of the emulated data fall as the number of initial bit flips increase. Second, the raw clustering follows the expected behavior in all the panels,  rising briefly before decaying down to incoherent uniform randomness at around $t\approx 12$. Third and most importantly however, we note that increasing the number of initial bit flips degrades the ability of our post-selection to correct for errors. While two initial bit flips (left panel) results in post-selected experimental clustering that follows the emulated curve relatively closely until $t\approx 4$ when it begins to degrade, three bit flips (middle panel) results in agreement out to only $t\approx 2$, and four initial bit flips (right panel) sees immediate disagreement between emulation and post-selected experiment after $t\approx 1$. In fact, for four initial bit flips, there is no QCA cycle for which post-selected experimental clustering is meaningfully improved over raw experimental clustering during the coherence window. This indicates that while our experimental protocol, including post-selection, is most appropriately applied at very low initial bit flip filling, it struggles to produce emergent complexity in the face of noise at higher filling fractions. This is likely because at higher initial isolated bit-flip filling, there are more error processes that cause amplitude renormalization but still keep the state vector in the protected sector of Hilbert space. That is, there are more locations in a higher-filling bit string where an erroneous bit flip will nonetheless result in domain wall conservation.

\end{document}